\documentclass[twocolumn]{aastex701}

\usepackage{placeins}

\newcommand{\rearth}{R_{\oplus}}
\newcommand{\mearth}{M_{\oplus}}
\newcommand{\rstar}{R_{\star}}
\newcommand{\mstar}{M_{\star}}
\newcommand{\fpp}{\mathrm{FPP}}
\newcommand{\nfpp}{\mathrm{NFPP}}
\newcommand{\triceratops}{\textsc{triceratops}}
\newcommand{\vespa}{\textsc{vespa}}
\newcommand{\blender}{\textsc{blender}}
\newcommand{\pastis}{\textsc{pastis}}

\newcommand{\toi}[1]{TOI-#1}

\shorttitle{High-resolution imaging in TESS validation}
\shortauthors{Collier et al.}
\accepted{to AAS Journals on August 20, 2026}

\begin{document}

\title{Assessing the Impact of High-Resolution Imaging on Statistical Validation of
TESS Planet Candidates}

\author[orcid=0009-0005-8024-2206, sname=Collier,gname=Michael]{Michael Collier}
\affiliation{Holaxis}
\email[show]{mike@holaxis.com}

\author[orcid=0000-0002-8965-3969,sname=Giacalone,gname=Steven]{Steven Giacalone}
\altaffiliation{NSF Astronomy and Astrophysics Postdoctoral Fellow}
\affiliation{Department of Astronomy, California Institute of Technology, Pasadena, CA 91125, USA}
\email{}

\author[orcid=0009-0008-5186-5326, sname=Derfer,gname=Brian]{Brian Derfer}
\affiliation{Holaxis}
\email{}


\author[orcid=0000-0002-5741-3047,sname=Ciardi,gname=David~R.]{David R. Ciardi}
\affiliation{NASA Exoplanet Science Institute, Caltech/IPAC, Pasadena, CA 91125, USA}
\email{}

\author[orcid=0000-0002-2361-5812,sname=Clark,gname=Catherine~A.]{Catherine A. Clark}
\affiliation{NASA Exoplanet Science Institute, Caltech/IPAC, Pasadena, CA 91125, USA}
\email{}

\author[orcid=0000-0002-1835-1891,sname=Crossfield,gname=Ian~J.~M.]{Ian J. M. Crossfield}
\affiliation{Department of Physics and Astronomy, University of Kansas, Lawrence, KS 66045, USA}
\email{}

\author[orcid=0009-0002-9833-0667,sname=Deveny,gname=Sarah]{Sarah Deveny}
\affiliation{Bay Area Environmental Research Institute, Moffett Field, CA 94035, USA}
\affiliation{NASA Ames Research Center, Moffett Field, CA 94035, USA}
\email{}

\author[orcid=0000-0001-9309-0102,sname=Fajardo-Acosta,gname=Sergio~B.]{Sergio B. Fajardo-Acosta}
\affiliation{Caltech/IPAC, Mail Code 100-22, Pasadena, CA 91125, USA}
\email{}

\author[orcid=0000-0001-9800-6248,sname=Furlan,gname=Elise]{Elise Furlan}
\affiliation{NASA Exoplanet Science Institute, Caltech/IPAC, Pasadena, CA 91125, USA}
\email{}

\author[orcid=0000-0002-0388-8004,sname=Gilbert,gname=Emily~A.]{Emily A. Gilbert}
\affiliation{NASA Exoplanet Science Institute, Caltech/IPAC, Pasadena, CA 91125, USA}
\email{}

\author[orcid=0000-0002-2532-2853,sname=Howell,gname=Steve~B.]{Steve B. Howell}
\affiliation{NASA Ames Research Center, Moffett Field, CA 94035, USA}
\email{}

\author[orcid=0000-0001-9811-568X,sname=Kraus,gname=Adam~L.]{Adam L. Kraus}
\affiliation{Department of Astronomy, The University of Texas at Austin, Austin, TX 78712, USA}
\email{}

\author[0000-0001-7233-7508]{Rachel A.~Matson}
\affiliation{U.S. Naval Observatory, Washington, D.C. 20392, USA}
\email{}

\author[0000-0001-5347-7062]{Joshua E. Schlieder}
\affiliation{Exoplanets and Stellar Astrophysics Laboratory (Code 667), NASA Goddard Space Flight Center, Greenbelt, MD 20771, USA}
\email{}

\author[orcid=0000-0003-2192-5371,sname=van~Eyken,gname=Julian~C.]{Julian C. van Eyken}
\affiliation{NASA Exoplanet Science Institute, Caltech/IPAC, Pasadena, CA 91125, USA}
\email{}





\author[0000-0002-2970-0532]{David Baker}  
\affiliation{Physics and Engineering Department, Austin College, Sherman, TX 75090, USA}
\email{}

\author[0000-0003-1464-9276]{Khalid Barkaoui}
\affiliation{Instituto de Astrof\'{i}sica de Canarias (IAC), E-38205 La Laguna, Tenerife, Spain} 
\affiliation{Astrobiology Research Unit, Université de Liège, 19C Allée du 6 Août, 4000 Liège, Belgium} 
\affiliation{Department of Earth, Atmospheric and Planetary Science, Massachusetts Institute of Technology, 77 Massachusetts Avenue, Cambridge, MA 02139, USA}
\email{}

\author[0000-0002-4746-0181]{\"Ozg\"ur Ba\c{s}t\"urk}
\affiliation{Ankara University, Faculty of Science, Astronomy \& Space Sciences Department, Tando\u{g}an, TR-06100, Ankara, T\"urkiye}
\email{}

\author[0009-0007-3012-1072]{Mario Basilicata}
\affiliation{INAF -- Brera Astronomical Observatory, Via E. Bianchi 46, 23807 Merate, Italy}
\email{}

\author[0000-0002-9436-2891]{Izuru Fukuda}
\email{izuru-fukuda@g.ecc.u-tokyo.ac.jp}
\affiliation{Department of Multi-Disciplinary Sciences, Graduate School of Arts and Sciences, The University of Tokyo, 3-8-1 Komaba, Meguro, Tokyo 153-8902, Japan}

\author[0000-0002-4909-5763]{Akihiko Fukui}
\email{afukui@g.ecc.u-tokyo.ac.jp}
\affiliation{Komaba Institute for Science, The University of Tokyo, 3-8-1 Komaba, Meguro, Tokyo 153-8902, Japan}
\affiliation{Instituto de Astrof\'{i}sica de Canarias (IAC), 38205 La Laguna, Tenerife, Spain}

\author[0000-0003-3986-0297]{Mourad Ghachoui}
\affiliation{Space Sciences, Technologies and Astrophysics Research (STAR) Institute, Université de Liège, 19C Allée du 6 Août, 4000 Liège, Belgium} 
\affiliation{Cadi Ayyad University, Oukaimeden Observatory, High Energy Physics, Astrophysics and Geoscience Laboratory, FSSM, Morocco}
\email{}

\author[0000-0003-1462-7739]{Micha{\"e}l Gillon}
\affiliation{Astrobiology Research Unit, Université de Liège, 19C Allée du 6 Août, 4000 Liège, Belgium}
\email{}

\author[0000-0001-8923-488X]{Emmanuel Jehin}
\affiliation{Space Sciences, Technologies and Astrophysics Research (STAR) Institute, Université de Liège, 19C Allée du 6 Août, 4000 Liège, Belgium}
\email{}

\author[0000-0002-6424-3410]{Jerome de Leon}
\email{jpdeleon@g.ecc.u-tokyo.ac.jp}
\affiliation{Komaba Institute for Science, The University of Tokyo, 3-8-1 Komaba, Meguro, Tokyo 153-8902, Japan}

\author[0000-0002-9428-8732]{Luigi Mancini}
\affiliation{Department of Physics, University of Rome ``Tor Vergata'', Via della Ricerca Scientifica 1, 00133 -- Rome, Italy}
\affiliation{INAF -- Turin Astrophysical Observatory, via Osservatorio 20, 10025 -- Pino Torinese, Italy}
\email{}

\author[0009-0003-1841-7381]{Francesca Manni}
\affiliation{Department of Physics, University of Rome ``Tor Vergata'', Via della Ricerca Scientifica 1, 00133 -- Rome, Italy}
\affiliation{INAF -- Turin Astrophysical Observatory, via Osservatorio 20, 10025 -- Pino Torinese, Italy}
\email{}

\author[0000-0001-7809-1457]{Gabriel Murawski}
\affiliation{Gabriel Murawski Private Observatory}
\email{}

\author[0000-0001-8879-7138]{Bob Massey} 
\affiliation{AAVSO}
\email{}

\author[0000-0001-9390-0988]{Luca Naponiello}
\affiliation{INAF -- Turin Astrophysical Observatory, via Osservatorio 20, 10025 -- Pino Torinese, Italy}
\email{}

\author[0000-0001-8511-2981]{Norio Narita}
\email{narita@g.ecc.u-tokyo.ac.jp}
\affiliation{Komaba Institute for Science, The University of Tokyo, 3-8-1 Komaba, Meguro, Tokyo 153-8902, Japan}
\affiliation{Astrobiology Center, 2-21-1 Osawa, Mitaka, Tokyo 181-8588, Japan}
\affiliation{Instituto de Astrof\'{i}sica de Canarias (IAC), 38205 La Laguna, Tenerife, Spain}

\author[0000-0001-6733-5380]{Justus Randolph}
\affiliation{American Association of Variable Star Observers}
\email{}

\author[0009-0009-0994-1767]{Roberto Zambelli}
\affiliation{Società Astronomica Lunae}
\email{}

\author[0000-0001-8227-1020]{Richard P. Schwarz} 
\affiliation{Center for Astrophysics \textbar \ Harvard \& Smithsonian, 60 Garden Street, Cambridge, MA 02138, USA}
\email{}

\author[0000-0002-0345-2147]{Abderahmane Soubkiou}
\affiliation{Astrobiology Research Unit, Université de Liège, 19C Allée du 6 Août, 4000 Liège, Belgium}
\email{}

\author{Gregor Srdoc}
\affiliation{Kotizarovci Observatory, Sarsoni 90, 51216 Viskovo, Croatia}
\email{}

\author[0000-0003-2163-1437]{Chris Stockdale}
\affiliation{Hazelwood Observatory, Australia}
\email{}

\author[0009-0007-2926-1924]{Toshi Suganuma}
\email{suganuma-toshi548@g.ecc.u-tokyo.ac.jp}
\affiliation{Department of Multi-Disciplinary Sciences, Graduate School of Arts and Sciences, The University of Tokyo, 3-8-1 Komaba, Meguro, Tokyo 153-8902, Japan}

\author[0000-0001-8665-4598]{Jiaqi Wang} 
\affiliation{National Astronomical Observatories, Chinese Academy of Sciences, 20A Datun Road, Chaoyang District, Beijing 100101, China}
\email{}

\author[0000-0003-2127-8952]{Francis P. Wilkin}  
\affiliation{Department of Physics and Astronomy, Union College, 807 Union St., Schenectady, NY 12308, USA}
\email{}

\author[0000-0002-5224-247X]{Sel\c{c}uk Yal\c{c}{\i}nkaya}  
\affiliation{Department of Astronomy \& Space Sciences, Faculty of Science, Ankara University, TR-06100, Ankara, T\"urkiye} 
\affiliation{Ankara University, Astronomy and Space Sciences Research and Application Center (Kreiken Observatory), Incek Blvd., TR-06837, Ahlatlıbel, Ankara, T\"urkiye} 
\affiliation{Astrobiology Research Unit, Universit\'e de Li\`ege, All\'ee du 6 Ao\^ut 19C, B-4000 Li\`ege, Belgium}
\email{}

\correspondingauthor{Michael Collier}

\begin{abstract}
High-resolution imaging is widely used to constrain false-positive
scenarios in exoplanet validation, but it is a finite follow-up
resource that reaches only a subset of candidates, and its
population-level impact on validation outcomes has not been quantified
through controlled removal experiments. Using an automated pipeline
built on \triceratops, we compute the false-positive probability
(FPP) of 443 TESS planet candidates. For the 264 planet
candidates with high-resolution imaging observations, we compute FPP
with and without the corresponding contrast curves, allowing us to
quantify the impact of the additional data. We find that 72\% of 68
contrast-curve-bearing validated planets would fail validation
without their adopted contrast curves. The fraction requiring imaging
decreases with increasing planet size, from $100\%$ below $1.7~\rearth$ to $33\%$
above $4~\rearth$: within our sample and \triceratops-based analysis,
the availability of high-resolution imaging directly limits the yield
of small-planet validation and the supply of validated targets for
atmospheric characterization. Our analysis statistically validates 64
new TESS planets with
sizes spanning $0.94$ to $7.83~\rearth$ across hosts of spectral type
M through F. Four of these are highly amenable to
JWST observations based on the transmission and emission spectroscopy
metrics, and each achieves validation only with its imaging
constraint.
\end{abstract}


\section{Introduction}\label{sec:intro}

The NASA {\it Kepler} mission transformed our
understanding of the exoplanet population through the identification of
more than 4000 planet candidates \citep{borucki2010, thompson2018}, many
of which have since been confirmed as bona fide planets and archived by
the NASA Exoplanet Archive \citep{akeson2013, christiansen2025}. The Transiting Exoplanet Survey Satellite
\citep[TESS;][]{ricker2015} was designed to build on this legacy by
surveying the nearest and brightest stars across nearly the entire sky. TESS identified
2241 planet candidates in its primary mission \citep{guerrero2021}, with NASA
mission-candidate holdings now exceeding 7000 and the TOI catalog continuing to
grow through the extended mission \citep{christiansen2025}. Many of these TESS planet candidates are ideal targets for
follow-up characterization with JWST \citep{hord2024}.

However, a transit-like signal does not by
itself guarantee the presence of a planet. Eclipsing binary stars (both those
orbiting the target and those blended into the photometric aperture from
nearby lines of sight) can produce signals that closely mimic planetary
transits. Instrumental artifacts, such as periodic signals induced by
spacecraft momentum dumps, can also masquerade as transits; this work
concerns the astrophysical false-positive scenarios. Before any planet
candidate can be considered reliable for demographic studies or prioritized
for costly follow-up observations, its astrophysical false-positive
probability should be assessed.

Over the years, a series of vetting tests have been developed to distinguish planet candidates from obvious astrophysical false positives, including searching
for centroid offsets \citep{bryson2013}, odd/even depth differences indicative
of eclipsing binaries \citep{coughlin2016}, and instrumental artifacts using
automated classifiers \citep{mccauliff2015, thompson2018, kostov2019}. These are vital for reducing the large number of periodic signals detected by transit surveys into a list of possible planets \citep{zink2020, kunimoto2025}, but additional analyses are required to promote these candidates to bona fide planets. In the absence of planet mass measurements, this is typically done using a probabilistic approach colloquially referred to as ``statistical validation.'' In short, statistical validation algorithms compute false-positive probabilities (FPPs) for planet candidates by comparing transiting planet and eclipsing binary models to the data, classifying them as true planets if their FPPs are sufficiently low.

Several frameworks have been developed for this purpose. \blender\
\citep{torres2004, torres2005, torres2011} validated landmark Kepler
discoveries through synthetic light-curve fitting but was too computationally
intensive for bulk application. \vespa\ \citep{morton2012, morton2015}
automated the process using Bayesian trapezoidal models and TRILEGAL
populations \citep{girardi2005}, enabling \citet{morton2016} to validate 1284
Kepler planets in a single study; \citet{rowe2014} validated 851 more in
multiplanet systems by leveraging the low false-positive rate of
multi-transiting candidates \citep{lissauer2012, lissauer2014}. \pastis\
\citep{diaz2014, santerne2015} provides a more rigorous likelihood treatment
at greater computational cost. More recent validation efforts also use
machine-learning classifiers to identify high-confidence planet candidates
from survey light curves and vetting metrics \citep{valizadegan2022,
valizadegan2023, valizadegan2025, lafarga2026}.

The transition to TESS sharpened the need for a validation approach that
explicitly accounts for crowding. TESS pixels span 21~arcsec, covering
roughly 28 times the sky area of a Kepler pixel, which means that blending
from unresolved neighbors is substantially more severe. At the same time,
Gaia \citep{gaiabrown2018, gaiadr3} and the TESS Input Catalog
\citep[TIC;][]{stassun2018, stassun2019} provided the stellar census
needed to model contamination from known nearby stars. A validation
procedure optimized for TESS must evaluate false-positive scenarios
involving this full population of known contaminants, and must be able to
incorporate observational constraints that limit the population of
\emph{unknown} contaminants.

\citet{giacalone2020, giacalone2021} developed \triceratops\ specifically to address this need. Unlike
its predecessors, \triceratops\ models transit-producing scenarios not only
for the target star but also for every nearby star that contributes flux to
the photometric aperture. For each star in the field, the tool evaluates a
suite of astrophysical configurations (including transiting planets,
eclipsing binaries, and background eclipsing binaries at various periods and
impact parameters) and computes scenario probabilities using a Bayesian
framework informed by the TESS light curve, known stellar properties from
the TIC, and priors drawn from empirical planet-occurrence rates and
binary-star statistics. The result is a pair of complementary metrics: the
FPP, defined as the overall probability that the signal is caused by something other than a planet transiting the target star (including false positives from both resolved and unresolved nearby stars), and the nearby false-positive probability (NFPP), which quantifies the likelihood that the signal originates from a resolved nearby star. More recently, \citet{gomezbarrientos2025} extended
\triceratops\ to incorporate multicolor transit photometry from ground-based
facilities (TRICERATOPS+), demonstrating that passband-specific likelihoods can further reduce the FPP for candidates
where single-band TESS data alone are insufficient. Since their inceptions, these tools have been used to validate dozens of TESS planets \citep{giacalone2022, giacalone2022b, mistry2023, mistry2023b, mistry2024, hord2024}. In even more recent years, related validation tools that leverage multi-color transit photometry \citep{pelaeztorres2024, jiang2026} have also been successfully used to verify the planetary natures of TESS signals.

Among the observational constraints that validation frameworks can ingest,
the high-resolution imaging contrast curve is one whose population-level
impact has not been measured through controlled experiments. Community
follow-up programs coordinated through the TESS Follow-up Observing Program
(TFOP) have obtained speckle, adaptive optics, and lucky imaging observations
for thousands of TOIs using facilities across the globe including the
Gemini, Keck, Palomar, and Lick Observatory telescopes
\citep[e.g.,][]{howell2021, ziegler2020, ziegler2021, schlieder2021}. These observations
produce contrast curves that place upper limits on the brightness of any
undetected companion as a function of angular separation from the target
star. When supplied to \triceratops, a contrast curve eliminates
false-positive scenarios that would require a blended source brighter than
the measured detection limit, directly constraining the population of
unknown contaminants that catalog data alone cannot reach.

The value of high-resolution imaging for individual targets is well
established \citep{howell2011, ciardi2015, furlan2017, ziegler2017}, but the population-level impact of contrast curves on
validation outcomes remains poorly quantified. Does the importance of imaging scale with planet
size, host star properties, or stellar field density? And can we quantify
which false-positive channels imaging suppresses, and by how much? If
imaging is the deciding factor for most small-planet validations, then
the allocation of high-resolution imaging time directly controls the yield
of the TESS validation enterprise, and the supply of statistically
validated targets for atmospheric characterization with JWST.

In this paper we apply an automated \triceratops-based validation
pipeline to 443 TESS planet candidates, compute the FPP of every
contrast-curve-bearing target both with and without its imaging, and
quantify the contribution of high-resolution imaging to validation
outcomes as a direct removal measurement. Across the sample we find
that imaging is decisive for the majority of small-planet validations;
we quantify the size dependence and its mechanism in
Section~\ref{sec:cc_impact}.

The layout of this paper is as follows. In
Section~\ref{sec:target-selection} we describe our target selection and
sample composition. In Section~\ref{sec:observations} we describe the TESS
photometry and the high-resolution imaging used in this work. In
Section~\ref{sec:methods} we describe our
ephemeris recovery, automated vetting pipeline, stellar-field assembly,
and FPP computation procedure. In Section~\ref{sec:results} we present
our results, including the validated planet catalog, the contrast-curve
removal analysis, and populations of interest (e.g., multi-planet systems). In Section~\ref{sec:discussion} we discuss
the implications of our findings for the TESS validation enterprise,
identify high-priority targets for JWST follow-up, compare our results
with previous work, and note methodological limitations. We summarize
in Section~\ref{sec:conclusions}.

\section{Target Selection}\label{sec:target-selection}

We drew our initial sample from the Exoplanet Follow-up Observing Program
(ExoFOP; \dataset[10.26134/ExoFOP5]{https::/dx.doi.org/10.26134/ExoFOP5}) catalog of TESS Objects of Interest
\citep[TOIs;][]{christiansen2025}. The majority of these TOIs were detected by the 
the TESS Science Processing Operations Center pipeline \citep{jenkins2016},
the MIT Quick Look Pipeline \citep{huang2020qlp, huang2020qlp2}, or the TESS Faint Star Search \citep{kunimoto2022faint}.
From the TOIs with a TESS Follow-up Observing Program Working Group
(TFOPWG) disposition of ``planet candidate'' (PC), we selected 603 that
fall within twelve project categories we defined for this work
(Section~\ref{sec:projects}). We excluded TOIs with existing
confirmed-planet (CP) or known-planet (KP) dispositions, retaining only
candidates awaiting validation.

For each TOI we queried ExoFOP to resolve its TIC identifier and
bootstrap ephemeris (orbital period, transit epoch, duration, depth)
\citep{christiansen2025}.

\subsection{Sample Composition}\label{sec:projects}

We defined 12 overlapping project categories, each targeting a region of
parameter space where statistical validation can address open questions
(Table~\ref{tab:projects}). Because the selection criteria overlap, 55
targets appear in more than one category, yielding 659 total memberships
across the 603 unique targets. The categories span host spectral types from
M to A and planet sizes from sub-Earth to sub-Saturn, and emphasize small
planets, the short-period Neptunian desert, and bright hosts amenable to
radial-velocity and atmospheric follow-up. We use these categories only to
describe how the heterogeneous input sample was assembled, not as
independent occurrence-rate or project-yield measurements.

\begin{deluxetable*}{lll}
\tabletypesize{\footnotesize}
\tablecaption{The 12 project categories, defined for this work, from which
  the 603 input TOIs were drawn. Counts are category memberships; because
  selection criteria overlap, 55 of the 603 unique TOIs belong to more than
  one category (659 total memberships). Selection used the ExoFOP catalog
  parameters at the time of sample assembly; final size classifications use
  our MCMC-derived radii (Section~\ref{sec:classification}).\label{tab:projects}}
\tablehead{\colhead{Category} & \colhead{Selection criteria} & \colhead{$N$}}
\startdata
M-dwarf sub-Earths & $T_\mathrm{eff} < 3900$~K, $R_p < 1.0~\rearth$ & 8 \\
M-dwarf super-Earths & $T_\mathrm{eff} < 3900$~K, $1.0$--$1.7~\rearth$ & 50 \\
M-dwarf sub-Neptunes & $T_\mathrm{eff} < 3900$~K, $1.7$--$2.6~\rearth$ & 52 \\
M-dwarf Neptunes & $T_\mathrm{eff} < 3900$~K, $2.6$--$4.0~\rearth$ & 22 \\
Neptunian desert (K dwarfs) & $3900$--$5200$~K, $3$--$8~\rearth$, $P < 4$~d & 73 \\
Neptunian desert (G dwarfs) & $5200$--$6000$~K, $3$--$8~\rearth$, $P < 4$~d & 113 \\
F-star sub-Neptunes & $6000$--$7500$~K, $1.7$--$2.6~\rearth$, $P > 3$~d & 47 \\
A-star small planets & $T_\mathrm{eff} > 7500$~K, $R_p < 4~\rearth$ & 10 \\
Ultra-short-period sub-Neptunes & $1.7$--$2.6~\rearth$, $P < 1$~d & 13 \\
Long-period planets & $P > 200$~d, $\geq 5$ TESS sectors & 29 \\
Bright small planets & $R_p < 2.6~\rearth$, $T_\mathrm{mag} < 10$ & 216 \\
Temperate bright Neptunes & $2.6$--$4.0~\rearth$, $T_\mathrm{mag} < 10$, $T_\mathrm{eq} = 200$--$500$~K & 26 \\
\enddata
\end{deluxetable*}

\section{Observations}\label{sec:observations}

\subsection{TESS Photometry}\label{sec:tess-photometry}

We retrieved all available SPOC 2-minute cadence light curves \citep{jenkins2016} for each
target using Lightkurve \citep{lightkurve2018}, which queries the
Mikulski Archive for Space Telescopes (MAST; \citealt{TESSdoi}; \dataset[10.17909/t9-nmc8-f686]{https://dx.doi.org/10.17909/t9-nmc8-f686}). When 2-minute data were
unavailable, we fell back to longer-cadence TESS-SPOC products
\citep{caldwell2020}. Targets
for which no SPOC light-curve products were available at any cadence were
excluded from further processing (12 targets). We deliberately excluded
QLP and other community pipeline products to maintain uniform systematics
treatment across the sample. We used the Pre-search Data Conditioning
Simple Aperture Photometry (PDCSAP) flux, which has been corrected for
instrumental systematics by the SPOC pipeline \citep{jenkins2016}. Downloaded sectors were
stitched into a single time series using the Lightkurve stitching
utilities, and NaN flux values were removed prior to detrending.

\subsection{High-Resolution Imaging}\label{sec:obs-imaging}

\begin{figure*}[ht!]
  \centering
  \includegraphics[width=\textwidth]{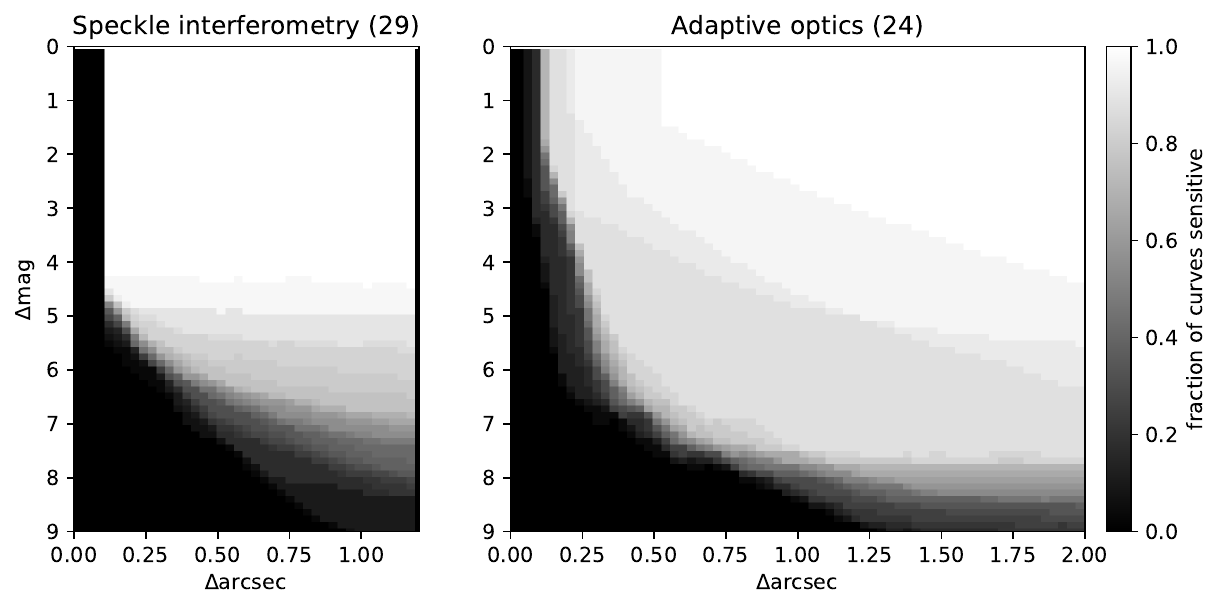}
  \caption{Sensitivity of the adopted contrast curves for the newly
    validated planets, separated by technique. Each pixel gives the fraction
    of curves sensitive to a companion of the given magnitude contrast
    ($\Delta$mag) at the given angular separation ($\Delta$arcsec), shading
    from white (all curves sensitive) to black (none sensitive). Speckle
    observations are confined to their $\sim 1.2''$ field of view, while the
    adaptive-optics curves extend to wider separations. Contrasts are in each
    observation's native band and are not directly comparable across
    instruments (cf.\ Table~\ref{tab:cc_sources}).
    \label{fig:cc_curves}}
\end{figure*}

The contrast curves used to constrain unresolved companions
(Section~\ref{sec:contrast-curves}) were drawn from high-resolution imaging
products archived on ExoFOP \citep{christiansen2025}, contributed by TESS
Follow-up Observing Program (TFOP) follow-up programs and collaborating
facilities. The adopted observations were obtained with the instruments
described below, and the sensitivity of the adopted curves is summarized in
Figure~\ref{fig:cc_curves}, with the specific observation adopted for each
validated planet (facility, instrument, band, and date) catalogued in
Table~\ref{tab:cc_sources}.

\subsubsection{Shane-3m/ShARCS}

The Shane Adaptive Optics infraRed Camera-Spectrograph (ShARCS; \citealt{mcgurk2014}), which is mounted behind ShaneAO \citep{kupke2012, gavel2014} on the Shane-3m Telescope at Lick Observatory, was used to collect AO imaging of TOIs analyzed in this paper. Observations were collected in natural guide star mode using a four-point dither pattern with a separation of $4\arcsec$ between each dither. Observations were collected with a Ks filter. Data were reduced using the publicly available \texttt{SImMER} pipeline \citep{savel2022}.\footnote{\url{https://github.com/arjunsavel/SImMER}} More information about observing strategy and the data can be found in Dressing et al. (submitted).

\subsubsection{Palomar-5m/PHARO}

The Palomar High Angular Resolution Observer (PHARO; \citealt{hayward2001}), which is mounted behind the natural guide star AO system P3K \citep{dekany2013} on the Palomar Observatory 5m Hale Telescope, was used to collect AO imaging of TOIs analyzed in this paper. Observations were collected using a five-point dither pattern with a separation of $5\arcsec$ between each dither and three exposures at each dither location. Images were primarily collected in the Kcont and Br$\gamma$ filters.

The AO data were processed and analyzed with a custom set of IDL tools with which the science frames were flat-fielded and sky-subtracted. Flat fields were generated from a median average of dark subtracted flats taken on-sky and normalized such that the median value of the flats is unity. Sky frames were generated from the median average of the dithered science frames. The reduced science frames were combined into a single combined image using an intra-pixel interpolation that conserves flux, shifts the individual dithered frames by the appropriate fractional pixels, and median-coadds the frames.

\subsubsection{Gemini-8m/Zorro and 'Alopeke}

Speckle interferometric observations were obtained for several TOIs using 'Alopeke and Zorro, dual-channel high-resolution imaging instruments mounted on the Gemini North and South 8m Telescopes, respectively \citep{scott2018}.\footnote{\url{https://www.gemini.edu/instrumentation/current-instruments/alopeke-zorro/}} Many thousands of 60 ms images were collected on two EMCCDs, each preceded by a narrow-band filter to minimize atmospheric dispersion. The full set of observations taken in 562 nm and 832 nm was then combined in Fourier space to produce their power spectrum and autocorrelation functions. From these, interferometric fringes were detected if a companion star was present within our $\sim 1 \farcs 2$ field of view, with an inner angle at the diffraction limit of the Gemini telescope. The data reduction pipeline produces final data products that include 5$\sigma$ contrast curves and reconstructed images \citep{horch1996, horch2012, howell2011}. The contrast curves at both 562 nm and 832 nm sample the spatial region near the target star from approximately 1 au to 50-100 au (depending on the distance to the target star) yielding contrast levels of 5-8 magnitudes. For more information about observing strategy and analysis procedure, see \cite{howell2011} and \cite{lester2021}.

\subsubsection{Gemini-8m/NIRI}

The AO-assisted near-infrared imager (NIRI) on the Gemini North 8m Telescope was used to observe some TOIs analyzed in this paper \citep{hodapp2003}. Images were collected with the Br$\gamma$ filter. The data were reduced using the Robo-AO reduction pipeline \citep{law2009, law2014}. More information about the observation and data reduction process can be found in \cite{law2014, baranec2016, ziegler2017}.

\subsubsection{Keck-10m/NIRC2}

The near-infrared camera instrument (NIRC2), which is mounted behind the natural guide star AO system on the Keck II Telescope \citep{wizinowich2000}, was used to observe several TOIs in this paper. Observations were acquired with a three-point dither pattern with a separation of $3\arcsec$ between moves, repeated three times for a total of nine exposures. Images were collected in several filters, including K, Kcont, Br$\gamma$, $J$cont, and $H$cont; the contrast curves adopted in our analysis (Section~\ref{sec:contrast-curves}) are those observed in K and Kcont. The data were processed and analyzed with the same software suite used for the Palomar PHARO observations. More information about observing procedure and analysis can be found in \cite{schlieder2021}.

\startlongtable
\begin{deluxetable*}{llllccc}
\tabletypesize{\scriptsize}
\tablecaption{Adopted contrast-curve provenance for the newly validated planets with a contrast curve. For each planet we list the single high-resolution imaging observation whose sensitivity curve entered the \triceratops\ false-positive-probability calculation: facility/instrument, band, UT date, and the achieved magnitude contrast $\Delta m$ at $0\farcs5$ and $1\farcs0$ read from the adopted curve. The CC-dep column marks planets that fail \triceratops\ validation without the contrast constraint; per-planet FPP with and without the contrast curve are listed in the validated-planet table and the machine-readable catalog. Provenance is the pipeline's own adopted-curve record. $\Delta m$ is the adopted curve's native-band contrast at the stated separation, as used by \triceratops; it is not directly comparable across instruments or bands (a star's optical-speckle vs NIR-AO limits, or its 562 and 832~nm speckle channels, can differ by $\sim$2~mag). The complete catalog, including the previously published planets, is provided as a machine-readable table.\label{tab:cc_sources}}
\tablehead{\colhead{Planet} & \colhead{Facility/Instrument} & \colhead{Band} & \colhead{UT Date} & \colhead{$\Delta m_{0\farcs5}$} & \colhead{$\Delta m_{1\farcs0}$} & \colhead{CC-dep}}
\startdata
TOI-213 b & Gemini-8m/Zorro and 'Alopeke & 562nm & 2020-11-29 & 4.33 & 4.44 & N \\
TOI-248 b & Gemini-8m/Zorro and 'Alopeke & 832nm & 2020-01-12 & 6.40 & 7.68 & Y \\
TOI-435 b & Gemini-8m/Zorro and 'Alopeke & 832nm & 2020-12-27 & 5.48 & 5.62 & N \\
TOI-783 b & Gemini-8m/Zorro and 'Alopeke & 832nm & 2023-04-07 & 6.85 & 7.81 & N \\
TOI-786 c & Gemini-8m/Zorro and 'Alopeke & 832nm & 2022-11-08 & 7.30 & 8.29 & Y \\
TOI-786 b & Gemini-8m/Zorro and 'Alopeke & 832nm & 2022-11-08 & 7.30 & 8.29 & N \\
TOI-789 b & Gemini-8m/Zorro and 'Alopeke & 832nm & 2020-12-26 & 6.25 & 6.76 & Y \\
TOI-797 b & Gemini-8m/Zorro and 'Alopeke & 832nm & 2020-03-12 & 6.41 & 7.21 & Y \\
TOI-873 b & Gemini-8m/Zorro and 'Alopeke & 832nm & 2020-11-29 & 5.78 & 6.02 & Y \\
TOI-1154 b & Gemini-8m/NIRI & Br$\gamma$ & 2019-11-08 & 6.95 & 7.90 & N \\
TOI-1245 b & Keck2-10m/NIRC2 & K & 2022-06-23 & 7.25 & 7.70 & Y \\
TOI-1262 b & Gemini-8m/Zorro and 'Alopeke & 832nm & 2020-02-17 & 6.20 & 6.89 & Y \\
TOI-1281 b & Shane-3m/ShARCS & Ks & 2021-03-29 & 3.04 & 4.60 & N \\
TOI-1432 b & Palomar-5m/PHARO & Br$\gamma$ & 2020-12-04 & 6.78 & 8.48 & Y \\
TOI-1435 b & Gemini-8m/Zorro and 'Alopeke & 832nm & 2020-06-07 & 6.60 & 7.75 & Y \\
TOI-1441 b & Palomar-5m/PHARO & Br$\gamma$ & 2021-08-24 & 7.13 & 8.16 & N \\
TOI-1806 c & Gemini-8m/Zorro and 'Alopeke & 562nm & 2021-02-03 & 4.92 & 4.98 & Y \\
TOI-1806 d & Gemini-8m/Zorro and 'Alopeke & 562nm & 2021-02-03 & 4.92 & 4.98 & Y \\
TOI-2238 b & Gemini-8m/Zorro and 'Alopeke & 832nm & 2020-12-23 & 6.27 & 7.19 & Y \\
TOI-2287 b & Gemini-8m/Zorro and 'Alopeke & 832nm & 2021-06-25 & 6.19 & 6.87 & Y \\
TOI-2293 b & Shane-3m/ShARCS & Ks & 2021-03-05 & 1.36 & 2.61 & Y \\
TOI-2392 b & Gemini-8m/Zorro and 'Alopeke & 832nm & 2022-10-09 & 6.37 & 6.99 & N \\
TOI-2441 b & Gemini-8m/Zorro and 'Alopeke & 832nm & 2022-01-13 & 6.58 & 7.32 & N \\
TOI-2540 b & Gemini-8m/Zorro and 'Alopeke & 832nm & 2022-02-12 & 6.53 & 7.52 & Y \\
TOI-3585 b & Palomar-5m/PHARO & Br$\gamma$ & 2021-08-08 & 7.23 & 7.96 & Y \\
TOI-4307 b & Gemini-8m/Zorro and 'Alopeke & 832nm & 2021-10-21 & 7.13 & 7.86 & Y \\
TOI-4324 b & Gemini-8m/Zorro and 'Alopeke & 832nm & 2021-02-25 & 6.92 & 7.98 & Y \\
TOI-4443 c & Palomar-5m/PHARO & Br$\gamma$ & 2022-05-19 & 6.91 & 8.37 & Y \\
TOI-4499 b & Palomar-5m/PHARO & Br$\gamma$ & 2022-05-19 & 6.97 & 8.37 & Y \\
TOI-4567 b & Gemini-8m/Zorro and 'Alopeke & 832nm & 2022-03-17 & 6.31 & 6.79 & Y \\
TOI-5165 b & Palomar-5m/PHARO & Br$\gamma$ & 2022-02-13 & 6.86 & 7.51 & Y \\
TOI-5177 b & Gemini-8m/Zorro and 'Alopeke & 832nm & 2025-02-16 & 7.15 & 9.06 & N \\
TOI-5289 b & Keck2-10m/NIRC2 & Kcont & 2023-08-05 & 7.36 & 7.56 & N \\
TOI-5493 b & Palomar-5m/PHARO & Kcont & 2023-12-02 & 6.57 & 7.91 & Y \\
TOI-5531 b & Palomar-5m/PHARO & Kcont & 2023-11-26 & 6.40 & 7.46 & Y \\
TOI-5543 b & Palomar-5m/PHARO & Br$\gamma$ & 2024-01-25 & 6.03 & 7.55 & Y \\
TOI-5696 b & Palomar-5m/PHARO & Kcont & 2024-02-17 & 6.80 & 8.19 & N \\
TOI-5730 b & Keck2-10m/NIRC2 & Kcont & 2023-06-10 & 7.32 & 7.69 & Y \\
TOI-5739 b & Palomar-5m/PHARO & Br$\gamma$ & 2023-06-07 & 3.72 & 4.83 & Y \\
TOI-5807 b & Palomar-5m/PHARO & Kcont & 2024-07-27 & 6.85 & 8.32 & Y \\
TOI-5947 b & Palomar-5m/PHARO & Kcont & 2024-08-08 & 6.92 & 7.54 & Y \\
TOI-5958 b & Palomar-5m/PHARO & Kcont & 2024-08-08 & 6.81 & 7.70 & Y \\
TOI-5961 b & Gemini-8m/Zorro and 'Alopeke & 832nm & 2023-07-04 & 6.55 & 7.27 & Y \\
TOI-5968 b & Palomar-5m/PHARO & Br$\gamma$ & 2023-08-29 & 7.03 & 8.31 & Y \\
TOI-5997 b & Gemini-8m/Zorro and 'Alopeke & 832nm & 2024-05-23 & 7.19 & 9.12 & Y \\
TOI-6098 b & Gemini-8m/Zorro and 'Alopeke & 832nm & 2025-01-08 & 6.91 & 8.05 & Y \\
TOI-6310 b & Keck2-10m/NIRC2 & Kcont & 2023-08-05 & 7.55 & 7.80 & N \\
TOI-6450 b & Gemini-8m/Zorro and 'Alopeke & 832nm & 2024-03-14 & 5.96 & 6.44 & N \\
TOI-6647 b & Palomar-5m/PHARO & Kcont & 2024-08-08 & 6.57 & 7.80 & N \\
TOI-6662 b & Gemini-8m/Zorro and 'Alopeke & 832nm & 2025-02-16 & 6.19 & 6.80 & Y \\
TOI-7182 b & Gemini-8m/Zorro and 'Alopeke & 832nm & 2025-08-09 & 7.16 & 9.23 & Y \\
TOI-7389 b & Palomar-5m/PHARO & Kcont & 2025-05-15 & 6.99 & 7.96 & Y \\
TOI-7391 b & Gemini-8m/Zorro and 'Alopeke & 832nm & 2025-08-07 & 5.35 & 5.52 & Y \\
\enddata
\end{deluxetable*}

\section{Methods}\label{sec:methods}

We developed an automated pipeline to compute false-positive probabilities
for a large sample of TESS planet candidates using the \triceratops\
framework \citep{giacalone2021}. The pipeline resolves each target,
recovers its transit ephemeris, applies a suite of vetting diagnostics,
assembles the inputs required for Bayesian validation, and executes
repeated Monte Carlo FPP evaluations. We describe each stage below.

\subsection{Ephemeris Recovery}\label{sec:ephemeris-recovery}

Rather than adopting the ExoFOP-reported ephemerides directly, we
independently recover each transit signal from the TESS photometry
(Section~\ref{sec:tess-photometry}) using a multi-stage procedure that
refines the period and epoch before validation.  19 TOIs fail this 
stage because they either have no reported orbital periods on ExoFOP
or because of persistent errors that prevented the steps described below
from completing. These TOIs are removed from our sample before beginning
the vetting steps in Section~\ref{sec:vetting}.

\subsubsection{Transit-Protected Detrending}

We detrended the stitched light curve using the Savitzky--Golay filter
implemented in Lightkurve, adopting a filter window several times the
transit duration. To prevent the filter from distorting the
transit signal, we first masked a window of $2 \times$ the transit
duration centered on each predicted transit midtime, excluding the
in-transit points from the filter fit; we refer to detrending with this
mask applied as ``transit-protected.'' In a second pass,
we recentered this mask on the BLS-recovered ephemeris
(Section~\ref{sec:bls}) and repeated the detrending.

\subsubsection{Two-Pass BLS Search}\label{sec:bls}

We searched for the transit signal using the Box Least Squares algorithm
\citep[BLS;][]{kovacs2002} in a two-pass configuration. The first pass
operates on the raw PDCSAP light curve, without the Savitzky--Golay
detrending applied, breaking
the circular dependency between transit masking and period recovery. The
second pass runs on the transit-protected detrended light curve.

Each BLS search explores a narrow period grid centered on the ExoFOP
bootstrap period ($\pm 0.1\%$). If the best-fit period rails against the
grid boundary, the search automatically widens to $\pm 0.5\%$. We {\bf also} check
for period aliases at factors of $1/2$, $2$, $1/3$, $3/2$, and $2/3$ of
the detected period; if the highest-power period occurs at an alias of the bootstrap period
or the widened search still rails, the pipeline falls back to
the bootstrap period.

We define the BLS signal-to-noise ratio as the ratio of the detected
depth to its formal uncertainty and adopt a detection threshold of
SNR $\geq 7$. BLS diagnostic products carried forward include the
best-fit period, epoch, duration, depth, odd/even transit depths with
uncertainties, and the phased secondary eclipse depth.

Targets where the BLS search failed to detect the signal above the SNR
threshold were excluded as ``not detected'' (21 targets).

\subsubsection{Global Transit MCMC}\label{sec:global-mcmc}

For targets with a BLS detection, we fit a transit model at the fixed BLS
period to the phase-folded, detrended light curve using Hamiltonian
Monte Carlo. The model uses the limb-darkened transit light curve from
\texttt{jaxoplanet} \citep{jaxoplanet} with orbital parameters from a \texttt{TransitOrbit}
model, sampled via the No-U-Turn Sampler \citep[NUTS;][]{hoffmangelman2014}
implemented in NumPyro \citep{phan2019} with JAX as the computational
backend \citep{jax2018}.

Before fitting, we trim the folded light curve to $\pm 3$ transit
durations, iteratively clip $5\sigma$ outliers using the median absolute
deviation (MAD) of model residuals, and bin to the longest sector cadence
present in the multi-sector stitch. The model has six free parameters:
the epoch offset ($t_0$), planet-to-star radius ratio ($R_p/\rstar$),
impact parameter ($b$), log transit duration ($\ln d$), and two quadratic
limb-darkening coefficients ($q_1, q_2$) in the uninformative
parameterization of \citet{kipping2013}.

We assess convergence using three criteria: (1) the effective sample size
for the epoch offset must be $\geq 100$, (2) the fraction of divergent
transitions must be $\leq 5\%$, and (3) the split $\hat{R}$ statistic
must be $\leq 1.03$ when multiple chains are available. The posterior
summaries provide the refined epoch, $R_p/\rstar$, impact parameter,
transit duration, and physical limb-darkening coefficients ($u_1$, $u_2$)
used in downstream vetting diagnostics.

Targets where the MCMC did not converge were classified as ``detected but
not recovered'' and excluded from further computation (20 targets total
across all convergence-failure modes).

For a minority of multi-sector targets, a small error in the fixed BLS
period accumulates across the baseline and smears the phase fold even
after a converged global fit. When the fold quality
$Q = \mathrm{depth}/\sigma_{\mathrm{in\text{-}transit}}$ falls below
$0.25$, we therefore refine the ephemeris from the timing of individual
transits, fitting a weighted linear ephemeris $T_n = T_0 + nP$ with
outlier rejection and rerunning the global MCMC on the sharper fold when
it improves on the BLS$+$MCMC result.\footnote{This procedure produced
phase-folded transits qualitatively equivalent to those from the full
time-series MCMC of Section~\ref{sec:juliet-emcee}.}

\subsection{Transit Vetting}\label{sec:vetting}

For the 584 TOIs that passed the ephemeris recovery stage, 
we apply a suite of diagnostic tests to each recovered signal before
permitting FPP computation. These tests are designed to identify
eclipsing binaries that could contaminate the candidate sample.

\subsubsection{Odd/Even Depth Test}

We test for depth differences between odd- and even-numbered events at
the candidate ephemeris. Such differences are the hallmark of an
eclipsing binary whose true orbital period is twice the detected one:
when the primary and secondary eclipses have comparable depths, the
transit search recovers the binary at half its true period, and the
fold interleaves primary and secondary eclipses as alternating odd-
and even-numbered events of different depths. An eclipsing binary at
the detected period itself is not flagged by this test, because every
folded event is then the same primary eclipse; such a binary may
instead reveal itself through a secondary eclipse elsewhere in phase,
which the secondary-eclipse test
(Section~\ref{sec:secondary-eclipse}) addresses.

We employ two independent methods. The first estimates odd- and
even-transit depths from the BLS model at the recovered ephemeris, using
formal white-noise uncertainties. The second measures transit depths
directly on the detrended time series by computing the median in-transit flux
relative to the out-of-transit baseline for each parity subset. This
robust method accounts for correlated (red) noise by estimating the
time-averaged uncertainty following the \citet{pont2006} formalism as
implemented by \citet{hartmanbakos2016}, which inflates the white-noise
estimate based on the measured red-noise power on the transit timescale.
The robust method requires at least 3 transits of each parity and at
least 10 in-transit cadences per parity to produce reliable depth and
uncertainty estimates.

When both methods are available, both must indicate a depth discrepancy
exceeding $3\sigma$ to flag the target. This ``both must agree''
criterion reduces spurious flags from isolated noise artifacts. When
the robust method cannot be computed (e.g., for long-period targets with
few observed transits), the pipeline falls back to BLS-only vetting with
a stricter threshold of $10\sigma$. This dual-threshold design reduces
the risk that the BLS-only path, which does not account for correlated
noise, excludes targets whose apparent odd/even differences are inflated by
red noise, while retaining sensitivity to high-confidence
eclipsing-binary signatures among sparsely observed systems.

The odd/even test removed 49 targets via the dual-method path and an
additional 21 targets via the BLS-only fallback, for a total of 70
likely eclipsing-binary signatures.

\subsubsection{Secondary Eclipse Test}\label{sec:secondary-eclipse}

We check whether a secondary eclipse is present in the BLS-phased light
curve. A target is excluded if the secondary depth exceeds $10\%$ of the
primary depth and the secondary signal-to-noise ratio exceeds $3$. This
test removed 18 additional targets with likely eclipsing-binary
secondary-eclipse signatures.

\subsubsection{V-Shape Flag}

We record a V-shape diagnostic for targets with grazing-eclipse-like
morphology, defined as impact parameter $> 0.9$ and $R_p/\rstar > 0.1$
based on the MCMC posteriors. This diagnostic is informational and is
not used as an exclusion criterion: a target is not excluded solely because
it satisfies the V-shape thresholds, and such targets proceed to
\triceratops\ evaluation unless another vetting test fails. No planets
in the final validated catalog meet this criterion.

\subsubsection{Vetting Summary}

In total, the vetting tests removed 88 of the 584 processed targets:
70 from odd/even tests (49 dual-method $+$ 21 BLS-only fallback) and 18
from secondary eclipse detection. Combined with the 21 non-detections,
12 targets lacking SPOC data, and 20 MCMC non-convergences, 141 of the
584 processed TOIs were excluded before FPP computation.

\subsection{Light Curve Preparation}

For targets that survive vetting, the \triceratops\ input is the PDCSAP
light curve prepared during ephemeris recovery
(Section~\ref{sec:ephemeris-recovery}): detrended, folded on the
recovered ephemeris, and clipped using the $5\sigma$ model-residual
procedure described above. This folded light curve is then trimmed to
$\pm 2$ transit durations and binned to approximately 200 phase bins
when more than 200 points are present.

We also download target pixel files (TPFs) to obtain the TESS aperture
geometry and World Coordinate System (WCS) solution, which are needed for
the stellar field assembly in Section~\ref{sec:stellar-field}.

\subsection{Stellar Field Assembly}\label{sec:stellar-field}

We use the \triceratops\ stellar-field assembly machinery to query
the TESS Input Catalog \citep[TIC;][]{stassun2019} around each target via
\texttt{astroquery} \citep{ginsburg2019}, retrieving nearby stars along
with their broadband magnitudes and stellar parameters. For each sector,
we use the TPF WCS solution and aperture mask to compute the pixel
position of each nearby star, its aperture flux contribution, and the
intrinsic eclipse or transit depth required to reproduce the observed
signal after dilution.

We supplement the TIC-based field with a synthetic background stellar
population generated by the TRILEGAL galactic model \citep{girardi2005}
at the target coordinates. This population provides the prior for
background false-positive scenarios (e.g., blended eclipsing binaries).

\subsection{TRICERATOPS}

We compute false-positive probabilities using the \triceratops\
statistical validation framework \citep{giacalone2021, giacalone2022}.
For each target, \triceratops\ evaluates a set of astrophysical scenarios
by computing the marginal likelihood of each scenario given the observed
light curve, stellar field, and prior information. 
The false-positive probability is then defined as
\begin{equation}
\mathrm{FPP}=1-\left(P_{\rm TP}+P_{\rm PTP}+P_{\rm DTP}\right),
\end{equation}
where $P_{\rm TP}$, $P_{\rm PTP}$, and $P_{\rm DTP}$ are the probabilities of the three scenarios involving a transiting planet around the target star.
The nearby false-positive probability (NFPP) is the posterior probability
summed over all scenarios in which the signal originates from a star
other than the target.

The transit and eclipse forward models use the analytic formulation of
\citet{mandelagol2002} as implemented in PyTransit
\citep{parviainen2015}. Priors on planet occurrence rates follow
\citet{fressin2013}, planet eccentricities follow the Beta distribution
parameterization of \citet{kipping2013b}, binary-star populations and
mass-ratio distributions follow \citet{moedistefano2017}, and empirical
stellar mass--radius--temperature relations follow \citet{torres2010}.

\subsection{FPP Computation}

\subsubsection{Monte Carlo Sampling}

\triceratops\ evaluates scenario likelihoods via Monte Carlo integration,
so a single evaluation carries sampling noise. Following
\citet{giacalone2021}, who likewise ran the calculation 20 times per
validated target to confirm that a validating FPP was not a stochastic
outlier, we execute 20 independent runs per target, each with a distinct
random seed and each drawing $10^6$ Monte Carlo samples. We report the
mean, standard deviation, minimum, and maximum FPP and NFPP across the
20 seeds rather than relying on a single evaluation.

\subsubsection{Ephemeris Consistency}\label{sec:ephemeris-consistency}

{ The FPP calculation takes the candidate ephemeris as an input, and
the ephemeris adopted at the validation stage (the BLS-based recovery
ephemeris of Section~\ref{sec:ephemeris-recovery}) differs slightly
from the final time-domain MCMC ephemeris of
Section~\ref{sec:juliet-emcee}. To confirm that no validation rests on
that choice, we recompute the FPP for every validated planet a second
time using its final MCMC period and epoch, holding all other inputs
fixed, and require a planet to satisfy the validation thresholds under
both ephemerides. The sample as a whole is
insensitive to the substitution: the median per-target ratio of the
MCMC-ephemeris FPP to the recovery-ephemeris FPP is $0.99$. Eight
targets, however, marginally exceed an FPP or NFPP threshold under
their MCMC ephemeris; we exclude these from the validated sample as
ephemeris-inconsistent (Section~\ref{sec:classification}).}

\subsubsection{Dual-Contrast Mode}

For targets with adopted contrast curves, we run each set of 20 seeds
twice: once incorporating the contrast curve as an observational
constraint, and once without it. This dual-contrast mode enables a
direct, per-target measurement of the contrast curve's impact on
validation outcomes. Targets without adopted contrast curves receive a
single evaluation without an imaging constraint. Of the 443 computed
targets, 264 had contrast curves available and were evaluated in
dual-contrast mode.

\subsubsection{Contrast Curve Resolution}\label{sec:contrast-curves}

Contrast curves encode the magnitude difference at which a companion
star can be excluded as a function of angular separation from the
target, as measured by high-resolution imaging (adaptive optics, speckle
interferometry, or lucky imaging). \triceratops\ uses these curves to
constrain the presence of unresolved companions near the target,
suppressing false-positive scenarios that require such a companion.
These include both background scenarios, in which a chance-aligned star
such as a background eclipsing binary is blended within the photometric
aperture, and bound-companion scenarios, in which the transit or eclipse
arises on an unresolved star physically associated with the target.
For the background scenarios, contrast curves reduce the prior probability 
by determining the number of background stars that could reasonably exist within 
the limits of the curve. \triceratops\ does this by first simulating a population 
of background stars within a 0.1 deg$^2$ area around the target using TRILEGAL. 
Next, using the contrast curve, it determines the maximum separation each simulated 
background star could exist from the target star while remaining undetected. 
The prior probability is then the total number of simulated background stars 
multiplied by the ratio of sky area within which the star can exist to the 
sky area from which all simulated stars were drawn (0.1 deg$^2$). For bound-companion
scenarios, contrast curves reduce the prior probability by determining the maximum
projected separation a hypothetical stellar companion could exist around the target star,
then integrating the orbital period prior distributions from \citet{moedistefano2017} out to that
separation to determine the probability of a stellar companion existing around the star
within those limits.

From the high-resolution imaging described in
Section~\ref{sec:obs-imaging}, we adopt a single contrast curve per target.
When multiple observations were available for a
single target, we selected among them by the compatibility of the
photometric band with the TESS bandpass, favoring red-optical, then
near-infrared, then blue-optical filters, and broke ties by the deepest
$\Delta$mag exclusion at $0.5''$ angular separation. Each adopted curve is tracked with
provenance metadata identifying the instrument, filter, and observation
date (Table~\ref{tab:cc_sources}).

\triceratops\ defines stellar flux relations for a canonical set of
filters (TESS, Vis, Kepler, $J$, $H$, $K$, $g$, $r$, $i$, and $z$) and
maps common imaging bands onto this set internally. A small number of
specialized speckle and adaptive-optics products use filters that fall
outside it, and we map each of these onto the nearest canonical filter.
The near-infrared narrowband and continuum filters (Br$\gamma$,
$H$-continuum, and $J$-continuum) are assigned to the broadband $K$,
$H$, and $J$ relations respectively, and the Robo-AO LP600 long-pass
filter is assigned to the TESS relation. This narrowband-to-broadband
substitution is an approximation, but each filter lies within the same
atmospheric window as its broadband counterpart, so the flux ratio of a
companion of a given mass differs only marginally.

\subsection{Classification}\label{sec:classification}

We classify each target according to its mean FPP and NFPP across the 20
Monte Carlo seeds, using the thresholds established by
\citet{giacalone2021}:
\begin{itemize}
\item \textbf{Validated planet (VP):} $\fpp < 0.015$ and $\nfpp < 0.001$
\item \textbf{Possible planet (PP):} $\fpp < 0.5$ and $\nfpp < 0.001$
      (but not meeting VP criteria)
\item \textbf{Likely nearby false positive (NFP):} $\nfpp \geq 0.1$
\item \textbf{Unclassified:} all remaining targets (including those in
      the gap $0.001 \leq \nfpp < 0.1$)
\end{itemize}

We emphasize that the NFP designation is a likely classification rather
than a confirmation: $\nfpp \geq 0.1$ marks a substantial posterior
probability that the signal arises on a nearby star rather than the target,
not a determination that the candidate is an astrophysical false positive.

We impose two additional quality cuts beyond the FPP thresholds. First,
we require the host star's Gaia DR3 \citep{gaiadr3} Renormalized Unit
Weight Error (RUWE) to satisfy $\mathrm{RUWE} \leq 1.4$
\citep{lindegren2018}. Elevated RUWE values indicate unresolved binarity
or astrometric excess noise that could compromise the single-star
assumption underlying \triceratops\ \citep{elbadry2021}. We query RUWE through the Gaia
archive via \texttt{astroquery}, cross-matching through the TIC's Gaia
source identifier.

Second, we cross-match each target against the ExoFOP stellar-companion
catalog \citep{christiansen2025}, which compiles companions from
high-resolution imaging or Gaia, and exclude any target with a cataloged
companion not matched to a star already present in the TIC field within
$1.5''$. Such companions violate the stellar field completeness assumed by
\triceratops\ and could produce false validations. Together, the RUWE
and companion cuts removed 13 targets that would otherwise meet the VP
thresholds. For these targets, \triceratops\ provides a facility to add
the resolved companion directly to the star list used in the NFPP
calculation; rather than apply this special-case treatment to the small
number affected here, we omit them and note that incorporating such
systems through this feature is a promising avenue for future work.

Planet radii reported in this work are derived from the MCMC-fitted
$R_p/\rstar$ posterior (Section~\ref{sec:juliet-emcee}) combined with
TIC v8.2 stellar radii \citep{stassun2019}, rather than the ExoFOP catalog values used for
target selection. For 16 of 80 VPs, the MCMC-derived radius places the
planet in a different size bin than the catalog radius.

Finally, beyond the automated thresholds we apply a conservative
manual-vetting step and remove a small number of otherwise-passing
targets from our final catalog. We remove targets for post-validation
data-quality reasons, including insufficient observed-transit coverage,
irreducibly ambiguous ephemerides (e.g., fewer than three observed
transits combined with a statistically significant depth mismatch that
cannot be resolved without additional observations), and period posteriors
that are bimodal or poorly constrained near the BLS-SNR detection floor.
We also set aside targets that are in the process of being validated or
characterized by independent teams. We treat those teams' analyses as the
primary discovery papers for these systems
and direct the reader to the relevant in-preparation publications; we
therefore do not include them in our validated count.  We additionally
exclude, as ephemeris-inconsistent, the eight targets that marginally fail
a validation threshold under their final MCMC ephemeris
(Section~\ref{sec:ephemeris-consistency}). These removals are
tracked separately from the automated FPP, RUWE, and companion cuts and
are not used to tune the validation thresholds.

\subsection{Time-Domain MCMC for Table~\ref{tab:validated} Posteriors}\label{sec:juliet-emcee}

To derive precise transit parameters and uncertainties for
Table~\ref{tab:validated}, we fit each candidate that satisfies the FPP,
RUWE, and companion cuts with a second, independent MCMC in the time
domain. A separate fit is required because the global phase-folded MCMC
of Section~\ref{sec:global-mcmc}, which drives the ephemeris recovery,
the vetting tests of Section~\ref{sec:vetting}, and the inputs to the
FPP classification of Section~\ref{sec:classification}, holds the BLS
period fixed and therefore cannot supply a period uncertainty. This
stage is applied only to the validated subset rather than to the full
computed sample. We use the \texttt{juliet}
framework \citep{espinoza2019juliet} with a \texttt{batman}
\citep{kreidberg2015} transit model and the \texttt{emcee}
affine-invariant ensemble sampler \citep{foremanmackey2013}. All transits
are fit simultaneously in their native (unfolded) time axis on
transit-windowed ($\pm 3 \times$ duration), Savitzky--Golay-detrended
PDCSAP light curves at the available SPOC/TESS-SPOC cadence. Initial
walker positions, and the Normal priors on period and epoch described
below, are centered on the BLS solution; the time-domain posteriors are
not conditioned on the phase-folded MCMC posterior.

The free parameters are as follows: the orbital period $P$, transit epoch $t_0$, the
\citet{espinoza2018} $(r_1, r_2)$ reparameterization of impact parameter
and $R_p/\rstar$, the stellar mean density $\rho_\star$, the
\citet{kipping2013} limb-darkening coefficients $q_1, q_2$, an additive
baseline offset, and a white-noise jitter term. Eccentricity is fixed to
zero. Priors are Normal on $P$ (width $0.01$~d) and $t_0$
(width $0.1$~d) centered on the BLS solution; Uniform on
$(r_1, r_2, q_1, q_2) \in [0,1]$; Normal on $\rho_\star$ with mean and
width propagated in quadrature from TIC v8.2 $\mstar$ and $\rstar$
(defaulting to $10\%$ in $\mstar$ and $6\%$ in $\rstar$ when TIC errors
are missing for two final validated planets); and LogUniform on the
jitter between $0.1$ and $1000$~ppm. Using $\rho_\star$ rather than
$a/\rstar$ directly propagates stellar-parameter uncertainty into the
transit geometry. We run 100 walkers for 30000 steps with a 5000-step
burn-in. Planet radii are computed post-hoc by multiplying the
$R_p/\rstar$ posterior by Gaussian draws from the TIC $\rstar$
distribution.

We apply a single quality cut to the time-domain fits: the $R_p/\rstar$
signal-to-noise ratio, defined as the posterior median divided by half
the $16$th--$84$th percentile width, must exceed $3$. One candidate
(\toi{4307.02}) fails this cut and is removed. Because period is a
free parameter in the time-domain fit, we report a period uncertainty in
Table~\ref{tab:validated} for every validated planet, whereas the
per-transit regression path of Section~\ref{sec:global-mcmc} supplied a
period uncertainty for only 10 of the raw VPs. All transit-parameter
posteriors quoted in Table~\ref{tab:validated} ($P$, $t_0$,
$R_p/\rstar$, $b$, $\rho_\star$, derived $a/\rstar$, and $R_p$) come
from these time-domain fits.

\section{Results}\label{sec:results}

We applied the pipeline described in Section~\ref{sec:methods} to 603
TESS Objects of Interest. Of these, 584 were processed by the pipeline:
443 received FPP computations and 141 were excluded by automated vetting tests
and data-quality cuts (70 odd/even eclipsing binary signatures, of
which 49 were identified by the dual-method test and 21 by the BLS-only
fallback; 18 secondary eclipse detections; 21 BLS non-detections; 12
targets lacking any SPOC light-curve products; and 20 MCMC
non-convergences). The remaining 19 could not be processed: 17 lacked
reported orbital periods on ExoFOP at the time of analysis and 2
encountered transient pipeline errors.

Of the 443 computed targets, we classify 133 as satisfying the VP
thresholds ($\fpp < 0.015$ and $\nfpp < 0.001$), 92 as possible planets
(PP; $\fpp < 0.5$, $\nfpp < 0.001$, not meeting VP criteria), 72 as
likely nearby false positives (NFP; $\nfpp \geq 0.1$), and 146 as unclassified.
We then apply additional quality cuts to the 133 raw VPs: 13 are excluded
for $\mathrm{RUWE} > 1.4$ or unresolved stellar companions
(Section~\ref{sec:classification}), yielding 120 planets satisfying our
VP $+$ companion cuts. 40 additional targets are removed from our
final catalog by the conservative manual-vetting step. One candidate
(\toi{4307.02}) is removed because it fails the $R_p/\rstar$ signal-to-noise cut and has a
non-physical impact-parameter posterior ($b > 1$).
Six targets (\toi{6871.01}, \toi{5744.01}, \toi{6333.01},
\toi{6668.01}, \toi{6026.01}, and \toi{6717.01}) are removed for data-quality reasons, including
insufficient observed transits and irreducibly ambiguous or poorly
constrained ephemerides. Two further candidates (\toi{4090.01},
\toi{6078.01}) satisfy the FPP threshold only marginally (their per-seed
FPP straddles the $\fpp < 0.015$ cutoff) and have not completed the
vetting applied to the validated sample; they are conservatively withheld
pending further follow-up. 23 targets are set aside because
they are in the process of being validated by independent teams and are
not included in our validated count. Finally, eight further targets
(\toi{4401.01}, \toi{1207.01}, \toi{1057.01}, \toi{5629.01},
\toi{4898.01}, \toi{1761.01}, \toi{797.02}, and \toi{7384.01}) pass
every validation gate under the recovered ephemeris but marginally
exceed the FPP or NFPP threshold when the FPP is recomputed with their
final MCMC ephemeris (Section~\ref{sec:ephemeris-consistency}); under
the strict both-ephemeris criterion we set them aside as
ephemeris-inconsistent: likely planets, but not validated here.
This yields 80 final validated planets.
Of these, 16 were already published and are recovered here; our final
catalog thus contains 64 newly validated planets.

\subsection{Sample Overview}

Table~\ref{tab:validated} presents the full 64-row newly validated
planet catalog, while Table~\ref{tab:published-overlaps} lists the 16
previously published planets and their contrast-curve removal
metrics. Both are provided as machine-readable tables. For each of the
64 newly validated planets we report the mean FPP across 20 Monte Carlo
seeds ($10^6$ Monte Carlo samples per seed), the contrast-curve-removed
FPP ($\fpp_\mathrm{noCC}$), and derived planetary parameters including
period, radius, equilibrium temperature, and host-star properties; for
the 16 previously published planets we report only the FPP and
$\fpp_\mathrm{noCC}$ recovery metrics, since their physical parameters
are given in their discovery papers.
Table~\ref{tab:gated} summarizes the 141 excluded targets and their
exclusion reasons.
Table~\ref{tab:cc_sources} lists, for each newly validated planet with a
contrast curve, the high-resolution imaging observation whose sensitivity
curve was adopted in its false-positive-probability calculation, with the
full catalog (including the previously published planets) provided
machine-readably.

\startlongtable
\begin{deluxetable*}{lcccccccc}
\tabletypesize{\scriptsize}
\tablecaption{The 64 newly validated planets. Central values and
  asymmetric $1\sigma$ uncertainties for $P$, $t_0$, $a/\rstar$,
  $R_p/\rstar$, and $R_p$ are from the \texttt{juliet}$+$\texttt{emcee}
  time-domain MCMC posteriors (Section~\ref{sec:juliet-emcee}). The full
  set of these 64 newly validated planets, with all derived parameters,
  is provided as a machine-readable table; the 16 previously published
  planets we recover are listed separately in
  Table~\ref{tab:published-overlaps} with recovery metrics
  only.\label{tab:validated}}
\tablehead{
  \colhead{Planet} &
  \colhead{$P$ (d)} &
  \colhead{$t_0$ (BTJD)} &
  \colhead{$R_p$ ($\rearth$)} &
  \colhead{$R_p/\rstar$} &
  \colhead{$a/\rstar$} &
  \colhead{$T_\mathrm{mag}$} &
  \colhead{FPP} &
  \colhead{$\fpp_\mathrm{noCC}$}
}
\startdata
TOI-1154 b & $21.745490^{+0.000017}_{-0.000016}$ & $1686.66973^{+0.00087}_{-0.00092}$ & $2.73^{+0.14}_{-0.14}$ & $0.0269^{+0.0004}_{-0.0004}$ & $35.62^{+0.83}_{-1.87}$ & $9.32$ & $4.4\times10^{-5}$ & $1.5\times10^{-2}$ \\
TOI-1245 b & $4.8204406^{+0.0000036}_{-0.0000038}$ & $1684.55049^{+0.00101}_{-0.00099}$ & $2.35^{+0.09}_{-0.08}$ & $0.0411^{+0.0009}_{-0.0009}$ & $18.54^{+0.59}_{-0.63}$ & $12.18$ & $1.7\times10^{-4}$ & $7.5\times10^{-2}$ \\
TOI-1262 b & $20.877835^{+0.000057}_{-0.000076}$ & $1693.90233^{+0.00390}_{-0.00272}$ & $2.39^{+0.15}_{-0.15}$ & $0.0245^{+0.0008}_{-0.0008}$ & $35.60^{+2.23}_{-2.44}$ & $9.60$ & $5.5\times10^{-4}$ & $2.8\times10^{-2}$ \\
TOI-1281 b & $6.390997^{+0.000012}_{-0.000011}$ & $1689.66096^{+0.00108}_{-0.00107}$ & $2.68^{+0.14}_{-0.14}$ & $0.0186^{+0.0004}_{-0.0005}$ & $11.32^{+0.24}_{-0.47}$ & $9.66$ & $9.9\times10^{-4}$ & $9.9\times10^{-3}$ \\
TOI-1432 b & $6.1094927^{+0.0000064}_{-0.0000067}$ & $1715.32102^{+0.00096}_{-0.00095}$ & $2.39^{+0.14}_{-0.14}$ & $0.0240^{+0.0007}_{-0.0007}$ & $15.61^{+0.95}_{-1.06}$ & $9.47$ & $5.3\times10^{-4}$ & $1.9\times10^{-2}$ \\
TOI-1435 b & $0.6858239^{+0.0002692}_{-0.0000012}$ & $1683.42529^{+0.00158}_{-0.00402}$ & $1.08^{+0.09}_{-0.09}$ & $0.0121^{+0.0007}_{-0.0007}$ & $3.70^{+0.19}_{-0.28}$ & $9.98$ & $2.0\times10^{-3}$ & $4.8\times10^{-2}$ \\
TOI-1441 b & $22.092958^{+0.000143}_{-0.000049}$ & $1695.40620^{+0.00273}_{-0.01151}$ & $2.82^{+0.16}_{-0.15}$ & $0.0208^{+0.0005}_{-0.0005}$ & $27.10^{+0.78}_{-1.29}$ & $9.48$ & $1.9\times10^{-5}$ & $3.8\times10^{-3}$ \\
TOI-1806 d & $8.196064^{+0.000050}_{-0.000044}$ & $1905.22781^{+0.00284}_{-0.00366}$ & $2.88^{+0.25}_{-0.24}$ & $0.0659^{+0.0053}_{-0.0052}$ & $31.31^{+1.03}_{-1.11}$ & $12.62$ & $9.1\times10^{-4}$ & $2.0\times10^{-2}$ \\
TOI-1806 c & $0.5018934^{+0.0000025}_{-0.0000031}$ & $1901.20602^{+0.00361}_{-0.00261}$ & $1.32^{+0.16}_{-0.15}$ & $0.0302^{+0.0036}_{-0.0032}$ & $4.86^{+0.16}_{-0.16}$ & $12.62$ & $1.3\times10^{-3}$ & $6.2\times10^{-2}$ \\
TOI-213 b & $23.519904^{+0.000013}_{-0.000011}$ & $1343.33581^{+0.00086}_{-0.00091}$ & $2.60^{+0.20}_{-0.20}$ & $0.0355^{+0.0006}_{-0.0006}$ & $48.91^{+3.19}_{-3.44}$ & $9.78$ & $3.7\times10^{-10}$ & $8.6\times10^{-3}$ \\
TOI-2238 b & $3.390463^{+0.000015}_{-0.000012}$ & $2965.34238^{+0.00294}_{-0.00357}$ & $2.60^{+0.16}_{-0.16}$ & $0.0167^{+0.0007}_{-0.0007}$ & $7.08^{+0.39}_{-0.44}$ & $10.72$ & $6.4\times10^{-3}$ & $3.1\times10^{-2}$ \\
TOI-2245 b & $3.7980908^{+0.0000052}_{-0.0000053}$ & $2039.75668^{+0.00143}_{-0.00171}$ & $4.89^{+0.25}_{-0.25}$ & $0.0296^{+0.0005}_{-0.0005}$ & $6.95^{+0.15}_{-0.33}$ & $10.84$ & $6.4\times10^{-4}$ & \nodata \\
TOI-2287 b & $11.812041^{+0.000039}_{-0.000043}$ & $1688.98074^{+0.00469}_{-0.00344}$ & $1.80^{+0.12}_{-0.12}$ & $0.0138^{+0.0007}_{-0.0006}$ & $19.04^{+1.08}_{-1.20}$ & $8.57$ & $8.7\times10^{-3}$ & $1.1\times10^{-1}$ \\
TOI-2293 b & $6.068314^{+0.000017}_{-0.000016}$ & $1847.83967^{+0.00228}_{-0.00253}$ & $2.08^{+0.10}_{-0.10}$ & $0.0382^{+0.0014}_{-0.0015}$ & $22.40^{+0.68}_{-0.76}$ & $11.81$ & $8.4\times10^{-3}$ & $3.7\times10^{-2}$ \\
TOI-2392 b & $9.169362^{+0.000064}_{-0.000065}$ & $2038.63402^{+0.00680}_{-0.00606}$ & $2.37^{+0.17}_{-0.17}$ & $0.0124^{+0.0006}_{-0.0006}$ & $11.38^{+0.58}_{-0.75}$ & $10.40$ & $8.9\times10^{-4}$ & $4.9\times10^{-3}$ \\
TOI-2426 b & $4.0812^{+0.0011}_{-0.0012}$ & $1412.89926^{+0.00392}_{-0.00361}$ & $2.36^{+0.20}_{-0.20}$ & $0.0211^{+0.0014}_{-0.0015}$ & $10.53^{+0.43}_{-0.59}$ & $9.97$ & $1.4\times10^{-2}$ & \nodata \\
TOI-2441 b & $12.890155^{+0.000015}_{-0.000015}$ & $1335.50286^{+0.00190}_{-0.00218}$ & $2.73^{+0.13}_{-0.13}$ & $0.0535^{+0.0018}_{-0.0019}$ & $38.58^{+1.10}_{-1.24}$ & $12.83$ & $1.4\times10^{-4}$ & $8.0\times10^{-3}$ \\
TOI-248 b & $5.9910723^{+0.0000031}_{-0.0000031}$ & $1354.83621^{+0.00094}_{-0.00097}$ & $2.37^{+0.12}_{-0.12}$ & $0.0194^{+0.0004}_{-0.0004}$ & $12.75^{+0.56}_{-0.79}$ & $8.46$ & $8.7\times10^{-5}$ & $2.3\times10^{-2}$ \\
TOI-2540 b & $12.718739^{+0.000029}_{-0.000031}$ & $1480.94373^{+0.00105}_{-0.00106}$ & $2.29^{+0.26}_{-0.25}$ & $0.0316^{+0.0016}_{-0.0013}$ & $30.71^{+2.45}_{-3.74}$ & $8.27$ & $8.8\times10^{-5}$ & $4.6\times10^{-2}$ \\
TOI-3585 b & $3.5410696^{+0.0000051}_{-0.0000052}$ & $2827.11230^{+0.00084}_{-0.00081}$ & $4.26^{+0.33}_{-0.33}$ & $0.0477^{+0.0015}_{-0.0013}$ & $11.22^{+0.72}_{-0.85}$ & $11.23$ & $1.7\times10^{-4}$ & $1.9\times10^{-2}$ \\
TOI-3851 b & $3.383445^{+0.000012}_{-0.000012}$ & $2745.60008^{+0.00203}_{-0.00140}$ & $4.76^{+0.27}_{-0.27}$ & $0.0356^{+0.0010}_{-0.0012}$ & $8.04^{+0.44}_{-0.50}$ & $11.18$ & $8.6\times10^{-3}$ & \nodata \\
TOI-4307 b & $32.696842^{+0.000069}_{-0.000058}$ & $1349.75725^{+0.00444}_{-0.00381}$ & $1.39^{+0.07}_{-0.07}$ & $0.0116^{+0.0003}_{-0.0003}$ & $41.13^{+2.28}_{-2.55}$ & $6.64$ & $1.2\times10^{-7}$ & $4.8\times10^{-2}$ \\
TOI-4310 b & $29.566602^{+0.000065}_{-0.000063}$ & $1365.18230^{+0.00264}_{-0.00266}$ & $2.97^{+0.31}_{-0.30}$ & $0.0376^{+0.0014}_{-0.0010}$ & $37.47^{+1.89}_{-5.24}$ & $9.50$ & $1.0\times10^{-2}$ & \nodata \\
TOI-4324 b & $6.2509291^{+0.0000060}_{-0.0000057}$ & $1519.00087^{+0.00124}_{-0.00160}$ & $1.23^{+0.06}_{-0.06}$ & $0.0285^{+0.0010}_{-0.0010}$ & $26.35^{+0.88}_{-0.94}$ & $9.75$ & $7.6\times10^{-9}$ & $1.3\times10^{-1}$ \\
TOI-435 b & $3.352919^{+0.000012}_{-0.000013}$ & $1411.84424^{+0.00169}_{-0.00194}$ & $4.03^{+0.49}_{-0.45}$ & $0.0548^{+0.0038}_{-0.0033}$ & $12.22^{+0.82}_{-1.08}$ & $12.94$ & $1.5\times10^{-3}$ & $1.2\times10^{-2}$ \\
TOI-4443 c & $10.308817^{+0.000022}_{-0.000024}$ & $2011.83071^{+0.00137}_{-0.00128}$ & $2.22^{+0.16}_{-0.14}$ & $0.0200^{+0.0011}_{-0.0009}$ & $20.08^{+1.19}_{-1.33}$ & $7.91$ & $3.8\times10^{-4}$ & $2.2\times10^{-2}$ \\
TOI-4499 b & $5.53888^{+0.00014}_{-0.00018}$ & $3313.10351^{+0.00480}_{-0.00409}$ & $2.14^{+0.16}_{-0.16}$ & $0.0180^{+0.0011}_{-0.0011}$ & $12.73^{+0.53}_{-0.69}$ & $10.13$ & $2.1\times10^{-3}$ & $2.6\times10^{-2}$ \\
TOI-4567 b & $23.542503^{+0.000056}_{-0.000085}$ & $1548.57930^{+0.00380}_{-0.00289}$ & $2.33^{+0.13}_{-0.13}$ & $0.0351^{+0.0016}_{-0.0017}$ & $47.99^{+1.46}_{-1.54}$ & $11.92$ & $6.8\times10^{-4}$ & $2.0\times10^{-2}$ \\
TOI-4724 b & $3.5761^{+0.0013}_{-0.0012}$ & $3668.10909^{+0.00482}_{-0.00532}$ & $3.20^{+0.28}_{-0.27}$ & $0.0318^{+0.0021}_{-0.0022}$ & $10.67^{+0.60}_{-0.68}$ & $11.26$ & $1.5\times10^{-2}$ & \nodata \\
TOI-4851 b & $21.609130^{+0.000089}_{-0.000067}$ & $1500.88556^{+0.00529}_{-0.00460}$ & $2.77^{+0.16}_{-0.16}$ & $0.0221^{+0.0008}_{-0.0008}$ & $24.97^{+1.00}_{-1.10}$ & $9.41$ & $1.1\times10^{-2}$ & \nodata \\
TOI-5165 b & $3.577048^{+0.000043}_{-0.000051}$ & $2528.51740^{+0.00545}_{-0.00730}$ & $3.32^{+0.60}_{-0.52}$ & $0.0350^{+0.0057}_{-0.0050}$ & $10.37^{+0.70}_{-0.80}$ & $12.54$ & $4.1\times10^{-3}$ & $2.8\times10^{-2}$ \\
TOI-5177 b & $2.208640^{+0.000027}_{-0.000027}$ & $2526.89248^{+0.00634}_{-0.00558}$ & $1.93^{+0.13}_{-0.13}$ & $0.0105^{+0.0005}_{-0.0005}$ & $4.04^{+0.27}_{-0.22}$ & $8.85$ & $2.3\times10^{-4}$ & $2.7\times10^{-3}$ \\
TOI-5289 b & $3.441056^{+0.000021}_{-0.000018}$ & $3209.44584^{+0.00173}_{-0.00163}$ & $4.38^{+0.31}_{-0.30}$ & $0.0428^{+0.0017}_{-0.0017}$ & $9.82^{+0.40}_{-0.53}$ & $11.89$ & $1.9\times10^{-4}$ & $4.8\times10^{-3}$ \\
TOI-5493 b & $14.56485^{+0.00013}_{-0.00017}$ & $2238.29388^{+0.01067}_{-0.00516}$ & $2.36^{+0.17}_{-0.17}$ & $0.0209^{+0.0011}_{-0.0012}$ & $24.26^{+1.32}_{-1.66}$ & $10.74$ & $2.9\times10^{-3}$ & $2.9\times10^{-2}$ \\
TOI-5531 b & $4.535285^{+0.000040}_{-0.000033}$ & $1873.58934^{+0.00580}_{-0.00717}$ & $1.46^{+0.12}_{-0.12}$ & $0.0122^{+0.0007}_{-0.0007}$ & $10.77^{+0.60}_{-0.71}$ & $8.57$ & $5.2\times10^{-4}$ & $3.0\times10^{-2}$ \\
TOI-5543 b & $3.5424576^{+0.0000051}_{-0.0000049}$ & $2474.85680^{+0.00075}_{-0.00075}$ & $3.78^{+0.25}_{-0.25}$ & $0.0406^{+0.0011}_{-0.0011}$ & $11.14^{+0.73}_{-0.81}$ & $10.66$ & $2.1\times10^{-3}$ & $3.3\times10^{-2}$ \\
TOI-5696 b & $32.43016^{+0.00012}_{-0.00015}$ & $2537.50485^{+0.00184}_{-0.00176}$ & $2.59^{+0.16}_{-0.16}$ & $0.0288^{+0.0009}_{-0.0009}$ & $51.32^{+2.64}_{-3.44}$ & $9.14$ & $2.7\times10^{-5}$ & $8.4\times10^{-3}$ \\
TOI-5730 b & $1.9271070^{+0.0000098}_{-0.0000059}$ & $1900.38990^{+0.00180}_{-0.00265}$ & $1.40^{+0.10}_{-0.10}$ & $0.0277^{+0.0017}_{-0.0018}$ & $10.81^{+0.35}_{-0.35}$ & $11.16$ & $2.5\times10^{-6}$ & $4.8\times10^{-2}$ \\
TOI-5739 b & $8.430524^{+0.000064}_{-0.000058}$ & $1930.75084^{+0.00413}_{-0.00305}$ & $2.33^{+0.13}_{-0.13}$ & $0.0144^{+0.0006}_{-0.0006}$ & $12.44^{+0.68}_{-0.74}$ & $7.14$ & $1.4\times10^{-4}$ & $1.9\times10^{-2}$ \\
TOI-5807 b & $14.241844^{+0.000087}_{-0.000092}$ & $2799.68334^{+0.00417}_{-0.00379}$ & $2.78^{+0.16}_{-0.14}$ & $0.0155^{+0.0006}_{-0.0005}$ & $17.30^{+1.02}_{-1.15}$ & $6.88$ & $3.5\times10^{-3}$ & $6.1\times10^{-2}$ \\
TOI-5947 b & $6.932019^{+0.000022}_{-0.000028}$ & $2829.01389^{+0.00166}_{-0.00160}$ & $2.32^{+0.16}_{-0.15}$ & $0.0246^{+0.0011}_{-0.0010}$ & $17.56^{+1.05}_{-1.18}$ & $8.59$ & $2.1\times10^{-3}$ & $4.4\times10^{-2}$ \\
TOI-5958 b & $2.3061827^{+0.0000099}_{-0.0000062}$ & $2825.78557^{+0.00129}_{-0.00161}$ & $1.89^{+0.15}_{-0.15}$ & $0.0180^{+0.0011}_{-0.0011}$ & $7.81^{+0.46}_{-0.51}$ & $9.52$ & $8.2\times10^{-3}$ & $7.0\times10^{-2}$ \\
TOI-5961 b & $1.6175105^{+0.0000028}_{-0.0000021}$ & $2447.78106^{+0.00108}_{-0.00161}$ & $1.38^{+0.16}_{-0.17}$ & $0.0180^{+0.0007}_{-0.0008}$ & $7.08^{+0.74}_{-0.92}$ & $9.04$ & $6.2\times10^{-7}$ & $8.5\times10^{-2}$ \\
TOI-5968 b & $9.215040^{+0.000164}_{-0.000076}$ & $2833.39863^{+0.00350}_{-0.00789}$ & $2.69^{+0.17}_{-0.17}$ & $0.0233^{+0.0009}_{-0.0009}$ & $17.04^{+1.01}_{-0.83}$ & $9.50$ & $6.4\times10^{-4}$ & $2.7\times10^{-2}$ \\
TOI-5997 b & $5.655007^{+0.000012}_{-0.000012}$ & $1985.15973^{+0.00237}_{-0.00215}$ & $1.45^{+0.14}_{-0.13}$ & $0.0189^{+0.0012}_{-0.0010}$ & $17.42^{+1.26}_{-1.43}$ & $9.26$ & $3.3\times10^{-3}$ & $2.5\times10^{-1}$ \\
TOI-6098 b & $2.72985^{+0.00013}_{-0.00012}$ & $3692.56030^{+0.00134}_{-0.00148}$ & $2.30^{+0.15}_{-0.14}$ & $0.0196^{+0.0009}_{-0.0007}$ & $7.62^{+0.43}_{-0.49}$ & $8.77$ & $2.3\times10^{-3}$ & $5.4\times10^{-2}$ \\
TOI-6174 b & $3.5247^{+0.0016}_{-0.0018}$ & $3559.94615^{+0.00999}_{-0.00750}$ & $7.83^{+0.62}_{-0.61}$ & $0.0584^{+0.0029}_{-0.0030}$ & $7.52^{+0.41}_{-0.45}$ & $13.25$ & $8.1\times10^{-3}$ & \nodata \\
TOI-6310 b & $3.352783^{+0.000026}_{-0.000023}$ & $3211.59567^{+0.00167}_{-0.00188}$ & $3.11^{+0.34}_{-0.33}$ & $0.0400^{+0.0018}_{-0.0018}$ & $10.66^{+0.57}_{-0.83}$ & $11.84$ & $9.4\times10^{-6}$ & $2.9\times10^{-3}$ \\
TOI-6434 b & $2.77791^{+0.00022}_{-0.00022}$ & $3692.74165^{+0.00123}_{-0.00121}$ & $4.61^{+0.19}_{-0.19}$ & $0.0644^{+0.0017}_{-0.0018}$ & $10.82^{+0.23}_{-0.28}$ & $12.28$ & $3.4\times10^{-3}$ & \nodata \\
TOI-6450 b & $3.60990^{+0.00041}_{-0.00040}$ & $3695.19315^{+0.00159}_{-0.00168}$ & $7.36^{+0.55}_{-0.54}$ & $0.0851^{+0.0023}_{-0.0025}$ & $11.52^{+0.43}_{-0.65}$ & $12.90$ & $2.6\times10^{-3}$ & $1.3\times10^{-2}$ \\
TOI-6647 b & $3.9992^{+0.0010}_{-0.0012}$ & $3587.00512^{+0.00395}_{-0.00315}$ & $3.97^{+0.24}_{-0.24}$ & $0.0240^{+0.0009}_{-0.0009}$ & $7.15^{+0.33}_{-0.46}$ & $9.66$ & $1.8\times10^{-4}$ & $2.1\times10^{-3}$ \\
TOI-6662 b & $0.293950130^{+0.000000052}_{-0.000000055}$ & $1816.18379^{+0.00024}_{-0.00023}$ & $0.94^{+0.03}_{-0.04}$ & $0.0260^{+0.0005}_{-0.0006}$ & $3.85^{+0.12}_{-0.14}$ & $10.89$ & $1.6\times10^{-11}$ & $2.6\times10^{-2}$ \\
TOI-6729 c & $7.837435^{+0.000015}_{-0.000012}$ & $1329.41854^{+0.00290}_{-0.00301}$ & $3.01^{+0.17}_{-0.17}$ & $0.0148^{+0.0005}_{-0.0005}$ & $9.71^{+0.59}_{-0.67}$ & $9.05$ & $9.3\times10^{-3}$ & \nodata \\
TOI-6729 b & $3.863828^{+0.000038}_{-0.000033}$ & $1327.19364^{+0.01255}_{-0.01160}$ & $1.61^{+0.20}_{-0.17}$ & $0.0079^{+0.0009}_{-0.0008}$ & $5.99^{+0.35}_{-0.40}$ & $9.05$ & $2.5\times10^{-3}$ & \nodata \\
TOI-7182 b & $14.71272^{+0.00023}_{-0.00027}$ & $2460.51925^{+0.01603}_{-0.01620}$ & $2.08^{+0.27}_{-0.18}$ & $0.0142^{+0.0018}_{-0.0011}$ & $19.95^{+1.09}_{-1.20}$ & $8.19$ & $2.4\times10^{-3}$ & $6.8\times10^{-2}$ \\
TOI-7389 b & $26.345368^{+0.000094}_{-0.000129}$ & $1709.19456^{+0.00367}_{-0.00446}$ & $1.35^{+0.09}_{-0.08}$ & $0.0106^{+0.0004}_{-0.0005}$ & $31.99^{+1.03}_{-1.45}$ & $8.87$ & $3.8\times10^{-4}$ & $1.8\times10^{-2}$ \\
TOI-7391 b & $7.4108814^{+0.0000099}_{-0.0000092}$ & $1685.97929^{+0.00178}_{-0.00188}$ & $1.58^{+0.08}_{-0.08}$ & $0.0349^{+0.0014}_{-0.0015}$ & $28.81^{+0.89}_{-0.98}$ & $12.73$ & $4.9\times10^{-5}$ & $3.7\times10^{-2}$ \\
TOI-7550 b & $4.984407^{+0.000018}_{-0.000025}$ & $1441.75314^{+0.00295}_{-0.00208}$ & $2.90^{+0.13}_{-0.13}$ & $0.0422^{+0.0013}_{-0.0014}$ & $14.65^{+0.35}_{-0.45}$ & $12.12$ & $6.2\times10^{-3}$ & \nodata \\
TOI-783 b & $8.1125118^{+0.0000056}_{-0.0000057}$ & $1573.88164^{+0.00121}_{-0.00120}$ & $2.47^{+0.23}_{-0.22}$ & $0.0339^{+0.0010}_{-0.0009}$ & $23.31^{+1.75}_{-1.98}$ & $9.92$ & $1.2\times10^{-6}$ & $1.2\times10^{-2}$ \\
TOI-786 b & $12.6694005^{+0.0000090}_{-0.0000104}$ & $1326.82317^{+0.00158}_{-0.00112}$ & $2.41^{+0.12}_{-0.12}$ & $0.0168^{+0.0003}_{-0.0003}$ & $18.09^{+1.03}_{-1.12}$ & $9.82$ & $7.6\times10^{-4}$ & $1.2\times10^{-2}$ \\
TOI-786 c & $38.554097^{+0.000096}_{-0.000118}$ & $1349.00168^{+0.00485}_{-0.00398}$ & $2.60^{+0.14}_{-0.14}$ & $0.0182^{+0.0005}_{-0.0005}$ & $37.74^{+2.13}_{-2.38}$ & $9.82$ & $2.1\times10^{-3}$ & $2.5\times10^{-2}$ \\
TOI-789 b & $5.4470318^{+0.0000045}_{-0.0000058}$ & $1329.10706^{+0.00154}_{-0.00115}$ & $1.08^{+0.05}_{-0.05}$ & $0.0266^{+0.0010}_{-0.0010}$ & $25.21^{+0.83}_{-0.92}$ & $11.98$ & $7.4\times10^{-4}$ & $3.8\times10^{-2}$ \\
TOI-797 b & $1.80079489^{+0.00000081}_{-0.00000092}$ & $1326.81948^{+0.00072}_{-0.00073}$ & $1.21^{+0.05}_{-0.06}$ & $0.0235^{+0.0007}_{-0.0008}$ & $10.27^{+0.34}_{-0.35}$ & $11.71$ & $3.7\times10^{-5}$ & $6.2\times10^{-2}$ \\
TOI-873 b & $5.9312268^{+0.0000083}_{-0.0000080}$ & $1326.22036^{+0.00218}_{-0.00235}$ & $1.65^{+0.09}_{-0.09}$ & $0.0290^{+0.0012}_{-0.0014}$ & $21.42^{+0.66}_{-0.71}$ & $12.12$ & $5.6\times10^{-4}$ & $4.4\times10^{-2}$ \\
\enddata
\tablecomments{Central values and asymmetric $1\sigma$ uncertainties for
  $P$, $t_0$, $a/\rstar$, $R_p/\rstar$, and $R_p$ come from the
  \texttt{juliet}$+$\texttt{emcee} time-domain MCMC posteriors
  (Section~\ref{sec:juliet-emcee}); the period uncertainty is reported
  asymmetrically because the period posterior is skewed for a small tail
  of targets. Transit epochs $t_0$ are in
  $\mathrm{BTJD} = \mathrm{BJD} - 2457000$. $R_p$
  additionally propagates the TIC $\rstar$ uncertainty. FPP and
  $\fpp_\mathrm{noCC}$ are means of 20 \triceratops\ Monte Carlo seeds
  with/without the contrast curve; \nodata\ indicates that no contrast
  curve was available. The machine-readable version of this table lists
  the 64 newly validated planets and additionally includes original TOI
  identifiers, planet designations, $b$, $\rho_\star$, $T_\mathrm{eff}$,
  $\rstar$, $\log g$, the symmetric mean period uncertainty $\sigma_P$,
  NFPP, number of sectors, and number of transits. The 16 previously
  published planets we recover are provided as a separate
  machine-readable table (Table~\ref{tab:published-overlaps}) with their
  FPP and $\fpp_\mathrm{noCC}$ only, since their adopted physical
  parameters are given in the discovery papers cited there.}
\end{deluxetable*}

\begin{deluxetable*}{lllccc}
\tabletypesize{\scriptsize}
\tablecaption{Previously published planets recovered in our final
  validated-planet sample. These planets are provided as a separate
  machine-readable table reporting recovery metrics only (FPP,
  $\fpp_\mathrm{noCC}$, and CC-dependence); their adopted physical
  parameters are available in the discovery papers cited here. They are
  not counted among the 64 newly validated planets in
  Table~\ref{tab:validated}.\label{tab:published-overlaps}}
\tablehead{
  \colhead{Planet} &
  \colhead{TOI} &
  \colhead{Published source} &
  \colhead{FPP} &
  \colhead{$\fpp_\mathrm{noCC}$} &
  \colhead{CC-dep}
}
\startdata
TOI-1011 b & TOI-1011.01 & \citet{brinkman2025} & $5.7\times10^{-5}$ & $1.9\times10^{-2}$ & Y \\
TOI-1732 b & TOI-1732.01 & \citet{lafarga2026} & $8.0\times10^{-4}$ & $9.8\times10^{-3}$ & N \\
TOI-2200 b & TOI-2200.01 & \citet{lafarga2026} & $8.8\times10^{-6}$ & \nodata & \nodata \\
TOI-2453 b & TOI-2453.01 & \citet{lafarga2026} & $4.2\times10^{-4}$ & $3.1\times10^{-2}$ & Y \\
TOI-4030 b & TOI-4030.01 & \citet{lafarga2026} & $1.3\times10^{-3}$ & $2.5\times10^{-2}$ & Y \\
HD 35843 c & TOI-4189.01 & \citet{hesse2025} & $1.1\times10^{-6}$ & $5.6\times10^{-3}$ & N \\
TOI-4364 b & TOI-4364.01 & \citet{distler2025} & $2.4\times10^{-4}$ & $8.9\times10^{-2}$ & Y \\
TOI-4552 b & TOI-4552.01 & \citet{srivastava2026} & $1.2\times10^{-4}$ & $4.7\times10^{-2}$ & Y \\
TOI-4616 b & TOI-4616.01 & \citet{zonglang2026} & $1.1\times10^{-5}$ & $4.3\times10^{-2}$ & Y \\
K2-405 b & TOI-5116.01 & \citet{christiansen2022} & $6.3\times10^{-4}$ & $9.2\times10^{-3}$ & N \\
TOI-5159 b & TOI-5159.01 & \citet{lafarga2026} & $5.8\times10^{-4}$ & $4.9\times10^{-2}$ & Y \\
TOI-5486 b & TOI-5486.01 & \citet{lafarga2026} & $8.3\times10^{-5}$ & $8.2\times10^{-3}$ & N \\
TOI-5489 b & TOI-5489.01 & \citet{gomezbarrientos2026} & $2.4\times10^{-4}$ & $4.4\times10^{-2}$ & Y \\
TOI-5489 c & TOI-5489.02 & \citet{gomezbarrientos2026} & $1.5\times10^{-2}$ & $7.3\times10^{-2}$ & Y \\
TOI-5716 b & TOI-5716.01 & \citet{gomezbarrientos2026} & $1.5\times10^{-4}$ & $9.3\times10^{-2}$ & Y \\
HD 50554 c & TOI-6965.01 & \citet{liu2026} & $1.3\times10^{-10}$ & $2.3\times10^{-2}$ & Y \\
\enddata
\tablecomments{FPP values are means of 20 \triceratops\ Monte Carlo seeds.
  $\fpp_\mathrm{noCC}$ is the same calculation after removing the adopted
  contrast curve; \nodata\ indicates that no contrast curve was available.
  CC-dep is ``Y'' when the contrast-curve-removed calculation is at or above
  the validation threshold $\fpp=0.015$ (11 of 16 previously
  published planets) and ``N'' otherwise.}
\end{deluxetable*}

\begin{deluxetable}{lll}
\tabletypesize{\scriptsize}
\tablecaption{Excluded targets: representative examples by exclusion reason.
  Summary counts and the full 141-row machine-readable listing are
  given in the notes below.\label{tab:gated}}
\tablehead{
  \colhead{TOI} & \colhead{TIC} & \colhead{Exclusion reason}
}
\startdata
TOI-1036.01  & 146172354  & odd/even (dual-method) \\
TOI-1670.03  & 441739020  & odd/even (BLS-only) \\
TOI-1752.02  & 287139872  & BLS non-detection \\
TOI-1434.01  & 138588540  & MCMC non-convergence \\
TOI-1727.01  & 241225337  & secondary eclipse \\
TOI-2590.01  & 234832821  & no SPOC light-curve \\
\enddata
\tablecomments{Of the 141 excluded targets, 70 fail the
  odd/even test (49 via the dual-method path and 21 via the
  BLS-only fallback), 21 are BLS non-detections, 20 fail
  MCMC convergence, 18 show a secondary eclipse, and 12 lack
  SPOC light-curve products. Each target is counted once, under the first
  exclusion it triggers. ``Dual-method''
  odd/even flags require both the BLS and robust per-transit depth tests to
  exceed $3\sigma$ significance; ``BLS-only'' flags originate from the BLS
  odd/even test alone. ``BLS non-detection'' indicates a BLS signal-to-noise
  below the threshold $7.0$ during ephemeris recovery. ``MCMC
  non-convergence'' targets were detected by BLS but did not yield converged
  phase-folded global transit-model posteriors during ephemeris recovery.
  The examples above are the alphabetically-first TOI in each category; the
  full 141-row listing is available as a machine-readable table.}
\end{deluxetable}

\begin{figure*}[ht!]
  \centering
  \includegraphics[width=0.9\textwidth]{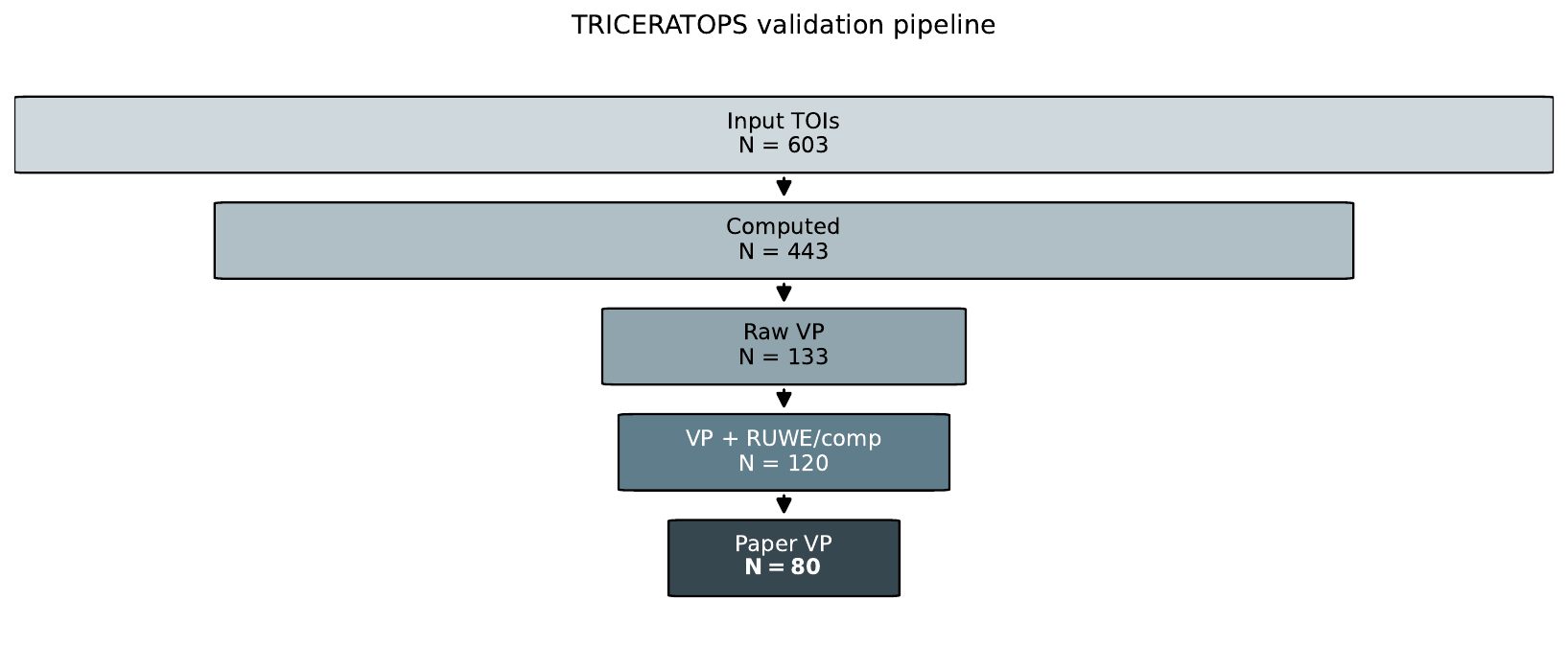}
  \caption{Schematic of the automated validation pipeline. Of the 603
    input TOIs, 584 are processed; 443 pass the vetting tests and
    receive full FPP computations. Of 133 raw validated planets (VPs)
    passing the FPP and NFPP thresholds,
    120 remain after the RUWE and companion cuts, and
    80 remain in our final catalog after the conservative
    manual-vetting step.
    \label{fig:pipeline}}
\end{figure*}

Figure~\ref{fig:pipeline} summarizes the pipeline flow from the initial
603-target input through transit vetting, FPP computation, and
classification. Of the 443 computed targets, 264 (60\%) have at
least one contrast curve ingested from ExoFOP, while 179 (40\%) lack
adopted contrast curves. The validation rate is higher among targets
with contrast curves ($68/264 = 25.8\%$) than among those without
($12/179 = 6.7\%$); we examine this difference, including potential
selection effects, in Section~\ref{sec:cc_impact}. Among the 80 VPs, the
median FPP is $6.1 \times 10^{-4}$ and 73 targets ($91\%$) achieve
perfect $20/20$ seed consistency, consistent with low sensitivity to
stochastic variation in the Monte Carlo draws.

The 80 VPs are drawn from the 12 overlapping project categories
described in Section~\ref{sec:projects}. Because category
membership overlaps by construction, we use these categories only as a
description of the input sample assembly rather than as independent
yield or occurrence-rate measurements. M-dwarf hosts account for 24 VPs
($30\%$ of the catalog), reflecting the favorable planet-to-star radius
ratios around cool dwarfs in transit surveys.

\subsection{Validated Planet Catalog}

\begin{figure}[ht!]
  \centering
  \includegraphics[width=\columnwidth]{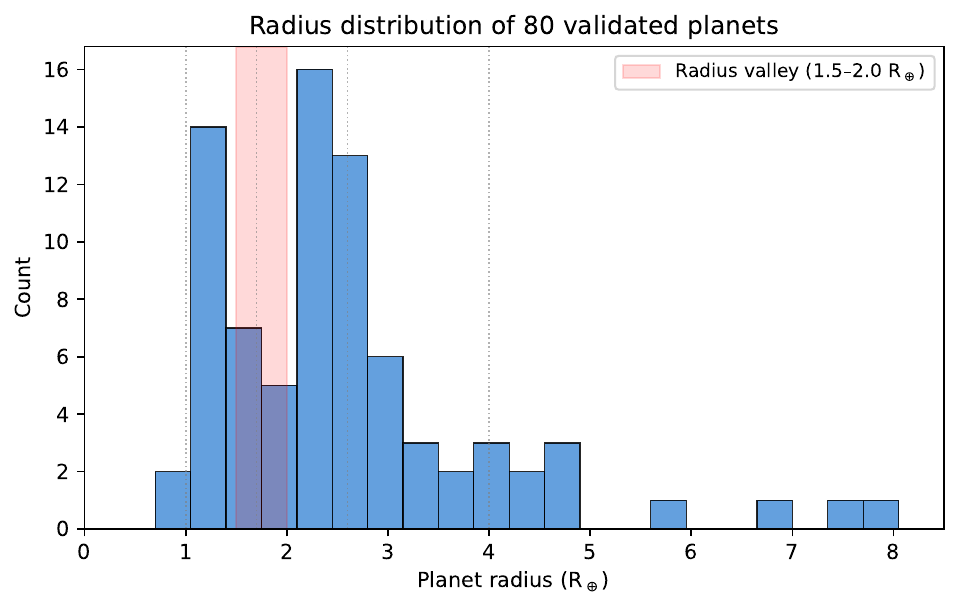}
  \caption{Radius distribution of the 80 validated planets. The
    shaded region near $1.7~\rearth$ denotes the radius valley, which
    overlaps with a local minimum in the distribution of our validated
    planets.
    \label{fig:radius_dist}}
\end{figure}

The 80 VPs span orbital periods from $0.29$~d (TOI-6662 b) to
$47.0$~d (HD 35843 c, TOI-4189.01) and radii from $0.94~\rearth$ (TOI-6662 b)
to $7.83~\rearth$ (TOI-6174 b). Planet radii are MCMC-derived
(Section~\ref{sec:juliet-emcee}); for 16 of 80 VPs, the MCMC radius
places the planet in a different size bin than the ExoFOP catalog value.
Figure~\ref{fig:radius_dist} shows the radius distribution, which
contains a dearth of planets near $1.7~\rearth$ resembling the radius valley
\citep{fulton2017, vaneylen2018}. The size-bin
breakdown is: 2 sub-Earths ($R_p < 1.0~\rearth$), 21 super-Earths
($1.0$--$1.7~\rearth$), 26 sub-Neptunes ($1.7$--$2.6~\rearth$), 20
Neptunes ($2.6$--$4.0~\rearth$), and 11 sub-Saturns
($4.0$--$8.0~\rearth$). No validated planets exceed the $8~\rearth$ classification
ceiling. The catalog spans all size bins, with the largest populations
in the sub-Neptune (26), super-Earth (21), and Neptune (20) bins, which is also the size
regime where, as we show in Section~\ref{sec:cc_impact}, contrast-curve
dependence is highest.

\begin{figure*}[ht!]
  \centering
  \includegraphics[width=\textwidth]{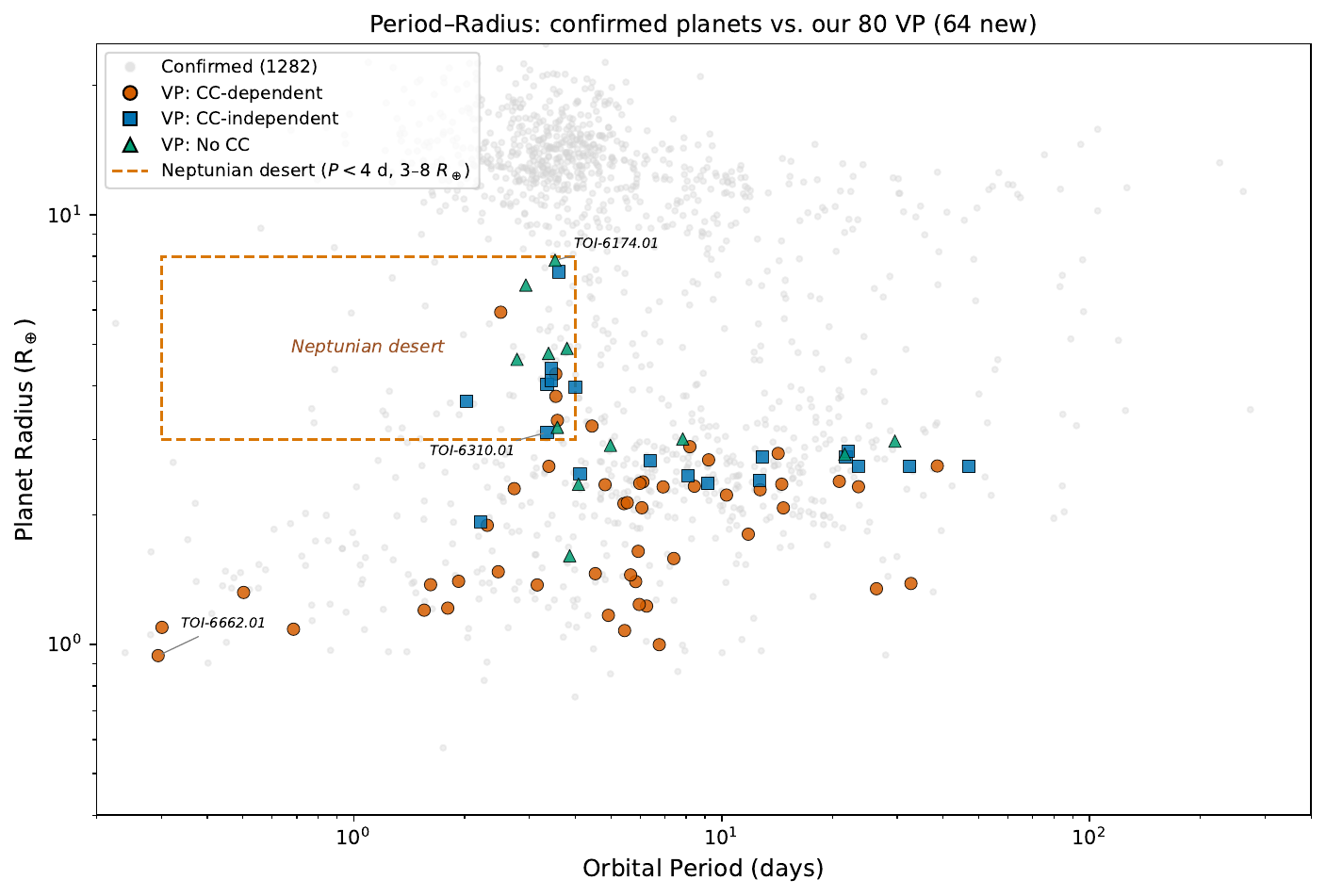}
  \caption{Period--radius distribution of our 80 validated planets
    compared with the TESS confirmed/known-planet population in the
    ExoFOP/TOI snapshot used for this analysis \citep{christiansen2025}
    (grey; $N=1282$ TOIs with
    TFOPWG disposition of Confirmed Planet or Known Planet). VP markers
    are coloured and shaped by their relationship to the contrast curve:
    CC-dependent (validated only with the curve, and would fall to
    possible-planet status without it), CC-independent (have a curve but
    remain validated without it), and No CC (no contrast curve available).
    The dashed rectangle marks
    the standard Neptunian desert boundary ($P < 4$~d,
    $3$--$8~\rearth$; see Section~\ref{sec:desert}). Labeled targets
    identify a lower-radius desert-boundary case (TOI-6310 b) and the
    largest retained VP (TOI-6174 b). We also label TOI-6662 b, which has the shortest period of all VPs in our sample. Twelve desert-dwelling
    candidates that passed our initial validation are set aside as
    overlaps with independent teams' in-progress validations
    (Section~\ref{sec:classification}); the retained desert sample is
    therefore substantially depleted and should not be read as
    representative.
    \label{fig:period_radius}}
\end{figure*}

The VP host stars span spectral types from mid-M
($T_\mathrm{eff} = 3244$~K; TOI-4616 b) to mid-F
($T_\mathrm{eff} = 6816$~K; TOI-5807 b). Figure~\ref{fig:period_radius}
shows the period--radius distribution of our VPs alongside the
confirmed/known TESS planet population from the ExoFOP/TOI snapshot used
for this analysis. Grouping hosts by stellar type and planet size shows
that super-Earths and sub-Neptunes together dominate the
contribution across all FGK and M spectral types, while we validate no
sub-Jupiters or Jupiters, consistent with the expectation that planets
above $\sim 8~\rearth$ typically require radial-velocity confirmation.
The breadth of the validated sample across spectral types reflects the
diversity of the input target categories described above.

\begin{figure*}[p!]
  \centering
  \includegraphics[page=1,width=\textwidth,height=0.88\textheight,keepaspectratio]{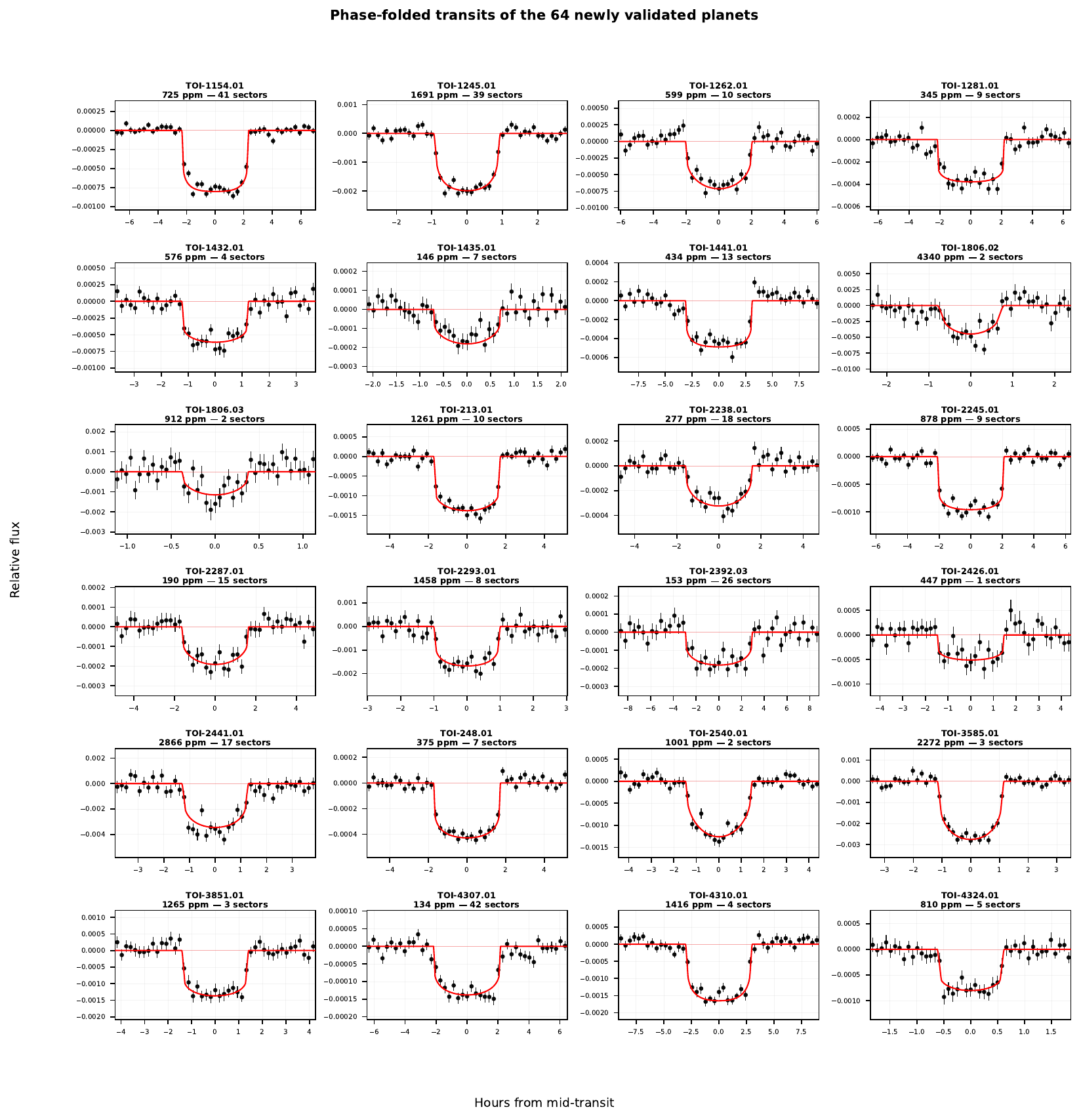}
  \caption{Phase-folded \texttt{juliet}$+$\texttt{emcee} best-fit
    transits for the 64 newly validated planets. Black markers are
    adaptively-binned photometry ($\sim 15$ bins across the transit
    duration, clipped to $2$--$30$~minutes) with $1\sigma$ error bars;
    red curves are the time-domain MCMC best-fit \texttt{batman} models
    evaluated at the posterior medians (Section~\ref{sec:juliet-emcee}).
    Each panel is clipped to $\pm 1.5$ transit durations from mid-transit.
    Panel titles report the target name, MCMC transit depth (ppm), and
    number of TESS sectors.
    \label{fig:transit_mosaic}}
\end{figure*}

\begin{figure*}[p!]
  \centering
  \includegraphics[page=2,width=\textwidth,height=0.88\textheight,keepaspectratio]{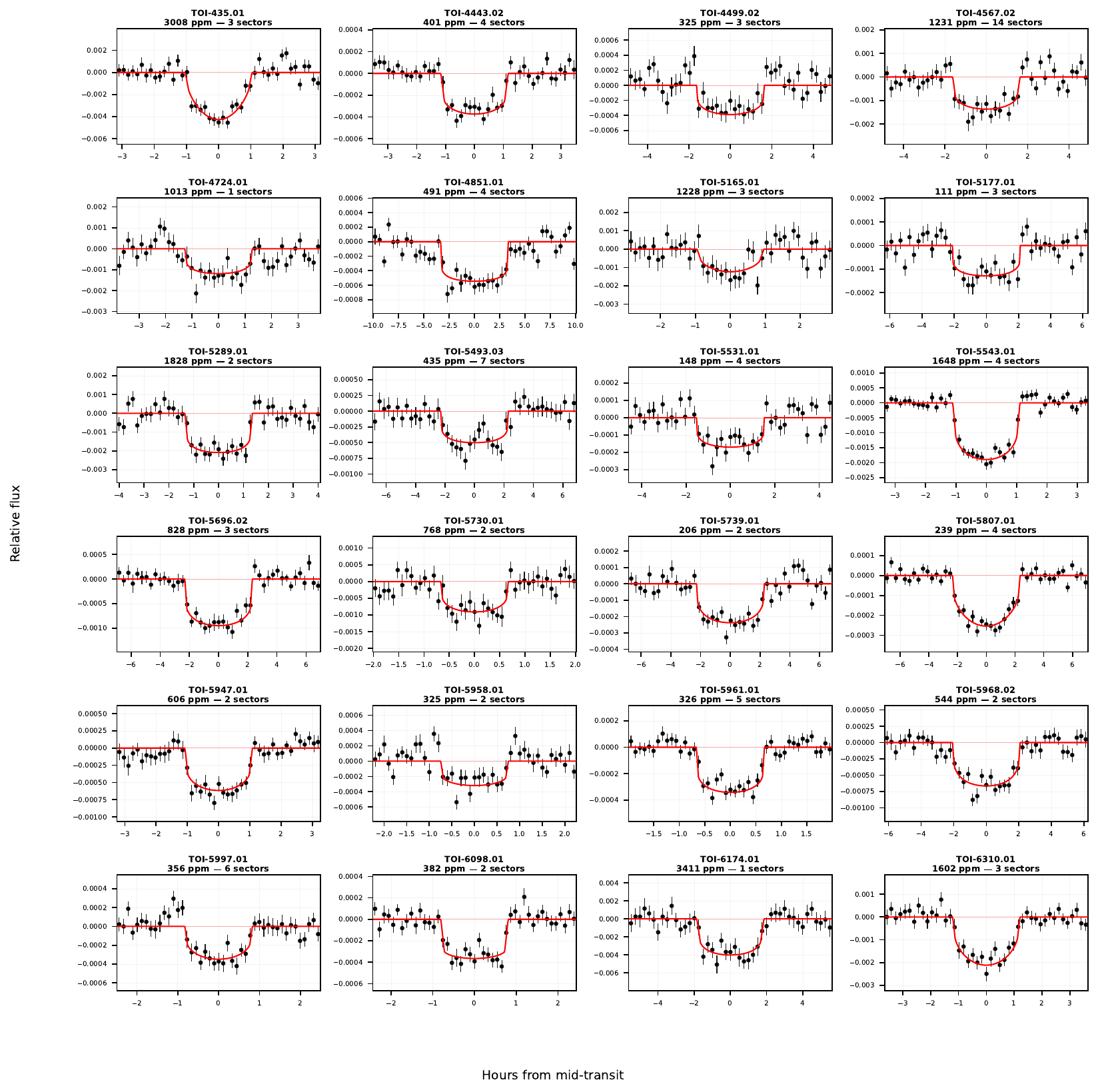}
  \par\smallskip\emph{Figure~\ref{fig:transit_mosaic} continued.}
\end{figure*}

\begin{figure*}[p!]
  \centering
  \includegraphics[page=3,width=\textwidth,height=0.88\textheight,keepaspectratio]{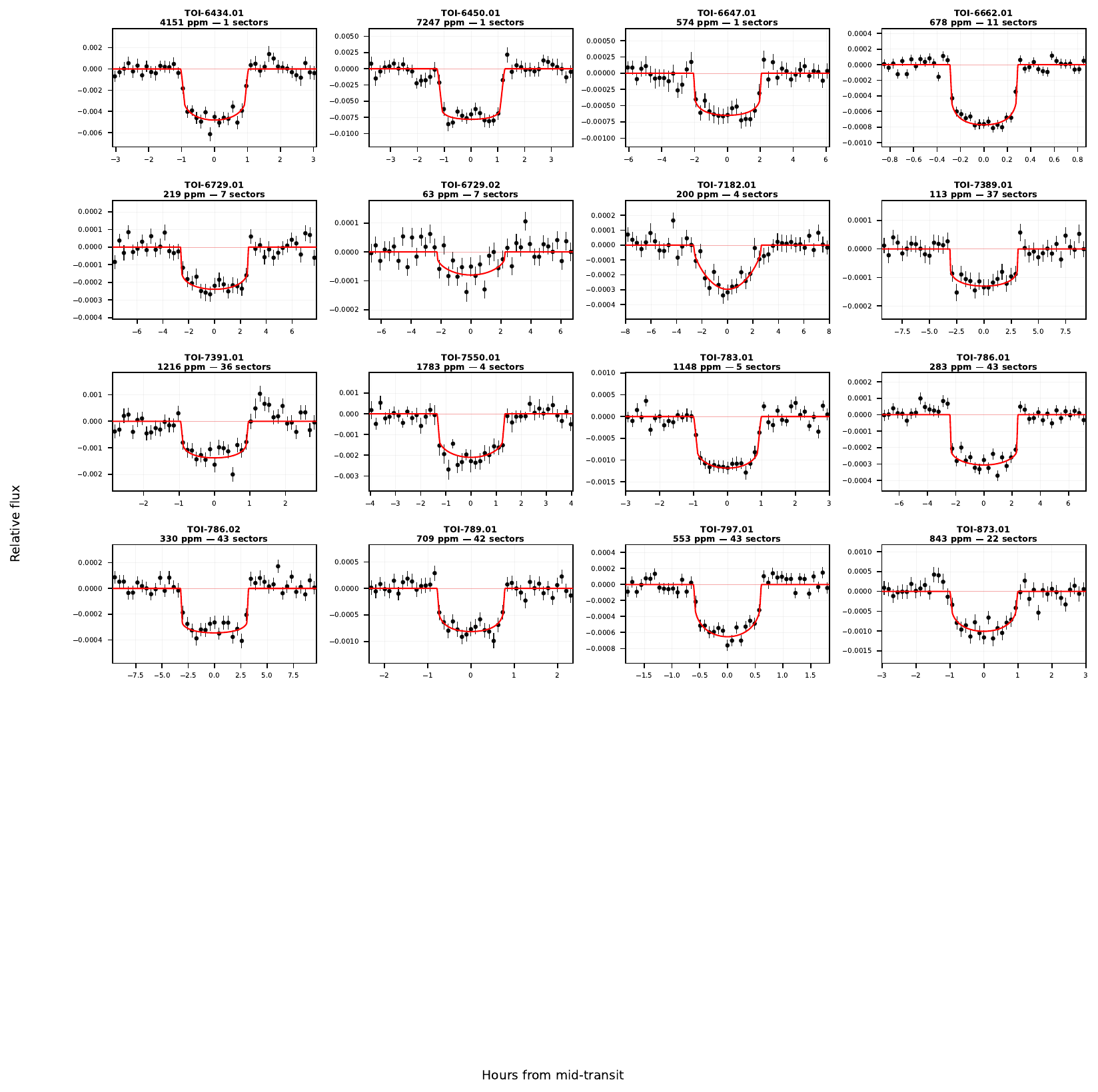}
  \par\smallskip\emph{Figure~\ref{fig:transit_mosaic} continued.}
\end{figure*}

\FloatBarrier

\subsection{Contrast Curve Impact}\label{sec:cc_impact}

Table~\ref{tab:cc_sources} gives, per newly validated planet with an
adopted contrast curve, the provenance of the single contrast curve adopted in its \triceratops\ FPP
calculation: the observing facility, instrument, band, UT date, and the
achieved $\Delta\mathrm{mag}$ at representative separations. The removal
results below are evaluated on the subset of validated planets and
computed targets for which a contrast-curve-removed FPP was available.

The central finding of this work is that, among our validated planets
with contrast curves, the majority of small planets require the imaging
constraint to achieve validation within the \triceratops\ framework. We
quantify this through a controlled removal experiment: for each of the
68 VPs that have at least one contrast curve, we recompute the FPP with
the contrast curve data removed ($\fpp_\mathrm{noCC}$) while holding
all other inputs fixed. A planet is ``contrast-curve-dependent''
(CC-dependent) if it would lose its VP classification without the
contrast curve (i.e.,
$\fpp_\mathrm{noCC} \geq 0.015$).

\begin{figure*}[ht!]
  \centering
  \includegraphics[width=0.9\textwidth]{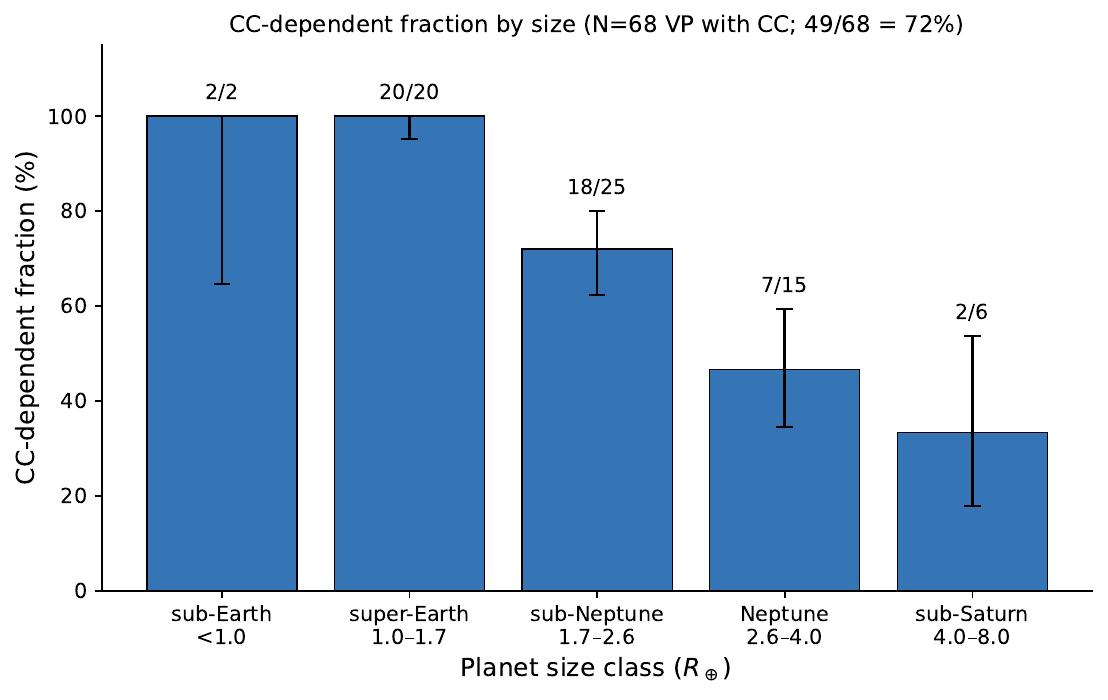}
  \caption{Contrast-curve-dependent fraction of the 68 (of 80)
    VPs that have adopted contrast curves, broken down by size bin; the
    12 VPs without archival imaging are omitted. The CC-dependence
    rises monotonically from $33\%$ (sub-Saturns, $4$--$8~\rearth$) to
    $100\%$ for super-Earths ($1.0$--$1.7~\rearth$) and sub-Earths.
    Bin-wise counts ($N_\mathrm{CC\mbox{-}dep}/N_\mathrm{VP}$)
    are annotated above each bar. Error bars show the $68.3\%$
    ($1\sigma$) uncertainty on each bin's fraction due to its small
    number of planets, computed as a Jeffreys binomial confidence
    interval \citep{cameron2011}; this asymmetric interval is appropriate
    for the modest per-bin counts (notably the sub-Saturns). The $100\%$
    bins (sub-Earths and super-Earths) are shown as lower limits.
    \label{fig:cc_dep}}
\end{figure*}

We find that 49 of 68 CC-bearing VPs (72\%) are CC-dependent: they would
fail the FPP threshold and drop to possible-planet (PP) status without
high-resolution imaging. None would drop to likely nearby false positive (NFP)
or false-positive status; all 49 transition exclusively from VP to PP.
Across the 264 CC-bearing computed targets (not restricted to VP),
$88.3\%$ are
FPP-sensitive to imaging constraints in the sense that their FPP
increases when contrast curves are removed; this is a weaker criterion
than CC-dependence, which requires loss of VP status. The overall
validation rate across the 443 computed targets is $18.1\%$ ($80/443$).
Restricting to the 264 targets with contrast curves, removing the
imaging constraint drops their validation rate from $25.8\%$ ($68/264$)
to $7.2\%$ ($19/264$), a factor of $3.6\times$ reduction. This result is
shown in Figure~\ref{fig:cc_dep}.

\subsubsection{Size Dependence}

The CC-dependence rate varies monotonically with planet size among the
VPs. Below $1.7~\rearth$, 22 of 22 VPs with contrast curves (100\%) are
CC-dependent; above $4.0~\rearth$, only 2 of 6 (33\%) require contrast
curves for validation. Figure~\ref{fig:cc_dep} displays the CC-dependence
rate by size bin, revealing a steep gradient: sub-Earths and
super-Earths are universally CC-dependent, while the rate
declines through the sub-Neptune and Neptune bins before reaching a
minimum for sub-Saturns. Across the full CC sample (not restricted to
VP), we measure a significant negative Spearman rank correlation between
planet radius and the logarithmic FPP reduction factor
($\rho = -0.37$, $p < 10^{-9}$, $N=264$), confirming that smaller planets are
systematically more reliant on imaging constraints.

The CC-dependence also varies with host-star spectral type. M dwarfs
show the highest VP loss rate without contrast curves
($19/23 = 83\%$), followed by F dwarfs ($9/12 = 75\%$), G dwarfs
($13/18 = 72\%$), and K dwarfs ($8/15 = 53\%$). The elevated M-dwarf
rate likely reflects both the smaller planet radii typical of M-dwarf
hosts and the more crowded stellar fields at fainter magnitudes.

This gradient has a clear physical interpretation. Small planets produce
shallow transits that are more easily reproduced by an eclipsing binary or
transiting planet on an unresolved companion at moderate flux ratio, and
contrast curves constrain these scenarios by limiting the brightness of any
companion within the photometric aperture. As the scenario decomposition
below shows, the suppressed probability is carried predominantly by
unresolved bound companions rather than chance-aligned background blends.
Larger planets, with their deeper transits, are intrinsically harder to
reproduce in this way, and a larger share of their false-positive
probability comes from channels that imaging constrains less directly.

\subsubsection{Scenario-Level Mechanism}\label{sec:mechanism}

Decomposing the false-positive probability that the imaging constraint
removes into the scenario classes of \citet{giacalone2021} shows that
contrast curves enable our validations chiefly by ruling out bound stellar
companions rather than background blends. For the 49
contrast-curve-dependent VPs, the bound-companion scenarios account for
essentially all of the false-positive probability the curve removes, with
the background and nearby-resolved channels together contributing less than
$1\%$ of it. A diagnostic restoration makes this concrete: holding every
other input fixed and replacing the with-curve probability of one scenario
class with its without-curve value, restoring the bound-companion scenarios
returns all 49 of these planets to non-validated status, whereas restoring
the background scenarios returns none. The bound-companion constraint alone
therefore reproduces the entire contrast-curve effect. Across the full set
of 264 contrast-curve targets, the bound-companion channel is the larger
false-positive channel for the great majority ($249$ of $264$) before any
imaging is applied, the background channel dominating only for a minority of
mostly larger planets.

\textit{Background channel.} Geometrically, the imaging constraint is most
effective against background blends: the number of chance-aligned background
stars within a given separation scales as the enclosed area,
$\propto \rho^{2}$, so tightening the admissible
radius from the $\sim 2.2''$ default of a \triceratops\ analysis to the
$\sim 0.3''$ of a typical contrast curve reduces the background population by
roughly $(2.2/0.3)^{2} \approx 50$. These reductions act on little
probability, however: across the 68 validated planets with a contrast curve
the summed background-blend probability is only of order $2 \times 10^{-4}$
before any imaging is applied. The curve removes most of it, but the
absolute change is small, which is why restoring the background scenarios
changes no validation.

\textit{Bound-companion channel.} The probability the contrast curves
actually remove is carried by bound companions, predominantly the scenario
of a transiting planet on an unresolved companion \citep{giacalone2021}.
Stellar multiplicity is common \citep{raghavan2010, moedistefano2017}, and unresolved
companions are a leading source of astrophysical false positives and a known
hazard for planet samples \citep{kraus2016}. The effect is visible even
among the candidates imaging fails to validate: of the 231 contrast-curve
targets that would exceed the FPP threshold without imaging, the
bound-companion scenarios are the dominant residual false positive for
$77\%$ with the curve removed but only $44\%$ with it applied, while the
nearby-resolved-star (NFPP) channel rises from $19\%$ to $41\%$. Because the
same targets are compared in both arms, this isolates the effect of the
imaging rather than a difference between samples: the curve markedly reduces
the bound channel's dominance and lifts to a comparable level the one
channel it cannot address, a known star already catalogued in the field but
blended in the TESS aperture.

The probability the curve removes from the bound channel also varies with
planet size. It falls by roughly a factor of two to three from the smallest
to the largest planets, and its share of the total FPP reduction declines
from nearly $100\%$ for small planets to about $70\%$ above $4~\rearth$ as
the background channel begins to contribute. Among the validated planets, by
contrast, the bound channel accounts for essentially all of the removed
probability at every size.

This imaging dependence extends to the planets most valuable for
atmospheric characterization: all 8 of our VPs that exceed the
\citet{kempton2018} transmission-spectroscopy thresholds recommended for
JWST follow-up (four of them newly validated) are CC-dependent, so
high-resolution imaging directly gates the supply of statistically
validated, JWST-accessible planets. We return to these targets, and the
emission-spectroscopy subset, in Section~\ref{sec:jwst}.

\subsection{Neptunian Desert Population}\label{sec:desert}

The Neptunian desert, the observed deficit of intermediate-size planets
($3$--$8~\rearth$) at short orbital periods
\citep[$P < 4$~d;][]{mazeh2016, owenlai2018, castrogonzalez2024}, provides a consistency
check on the contrast curve finding: these are large planets with deep
transits, exactly the regime where imaging should matter least. Indeed,
of the 11 desert VPs that have contrast curves, only 4 ($36\%$) are
CC-dependent, compared with $49/68$ ($72\%$) overall,
supporting the interpretation that the CC effect reported above is
driven by transit depth rather than by an artifact of our pipeline. Populating this region with validated planets also
adds objects in a regime relevant to atmospheric mass loss via photoevaporation and core-powered mass loss
\citep{owenwu2017, guptaschlichting2019, owenlai2018} and stripping via Roche-lobe overflow \citep{valsecchi2015, ginzburg2017, konigl2017, vissapragada2025}.

We validate 17 planets (13 newly validated) within the standard desert
boundaries ($P < 4$~d, $3$--$8~\rearth$), spanning $3.11~\rearth$
(TOI-6310 b) to $7.83~\rearth$ (TOI-6174 b); these appear within the
dashed desert-boundary overlay of Figure~\ref{fig:period_radius}.
A further twelve desert candidates that passed our initial validation
are set aside here as overlaps with independent teams' in-progress
validations (Section~\ref{sec:classification}). Because these set-asides
are concentrated in the desert, the retained desert sample is
substantially depleted and non-representative; we report it as a set of
individually validated planets rather than as a basis for demographic or
occurrence-rate inference about the desert.

\subsection{Multi-Planet Systems}\label{sec:multi_planet}

\begin{deluxetable*}{lccccccl}
\tabletypesize{\scriptsize}
\tablecaption{Multi-planet systems hosting two or more validated
  planets (4 systems, 8 VP; 3 systems entirely new).
  \label{tab:multi}}
\tablehead{
  \colhead{System} &
  \colhead{$N$ VP} &
  \colhead{Validated planets} &
  \colhead{$P_\mathrm{min}$ (d)} &
  \colhead{$P_\mathrm{max}$ (d)} &
  \colhead{Period ratio} &
  \colhead{New?} &
  \colhead{Notes}
}
\startdata
TOI-1806\tablenotemark{a}  & 2 & c (.03), d (.02)   &  0.502 &  8.196 &  16.33 & Y, Y & c USP; widely spaced \\
TOI-5489  & 2 & b (.01), c (.02)   &  3.152 &  4.921 &   1.561 & N, N & 4.1\% from 3:2 \\
TOI-6729  & 2 & b (.02), c (.01)   &  3.864 &  7.837 &   2.028 & Y, Y & 1.4\% from 2:1 (near-resonant) \\
TOI-786   & 2 & b (.01), c (.02)   & 12.669 & 38.554 &   3.043 & Y, Y & 1.4\% from 3:1 (near-resonant) \\
\enddata
\tablenotetext{a}{TOI-1806.01 is a previously confirmed planet at
$P = 15.15$~d (outside the VP sample validated here); the full
three-planet architecture therefore spans a period ratio of $30.2$
between the confirmed outer planet and our innermost validated planet,
while our two newly validated planets alone (c and d) span the
ratio of $16.33$ quoted in the table. We letter the two planets
validated here c and d in period order, leaving b available for the
earlier-confirmed (currently unlettered) TOI-1806.01.}
\tablecomments{Systems are ordered by $P_\mathrm{min}$. Periods are
  central values from the \texttt{juliet}$+$\texttt{emcee} MCMC
  posteriors (Section~\ref{sec:juliet-emcee}); uncertainties are
  given in the machine-readable VP catalog. ``$P_\mathrm{min}$'' and
  ``$P_\mathrm{max}$'' denote the shortest and longest validated
  orbital periods in each system. Period ratio is
  $P_\mathrm{max}/P_\mathrm{min}$. The ``New?'' column marks each
  planet in $P$-order as Y (not previously published) or N
  (previously published in a prior statistical validation paper). All
  4~systems yield 2 VP each in the final 80-VP catalog; 3 of the 4
  systems are entirely new, and TOI-5489 is the sole system where
  both planets were previously published \citep{gomezbarrientos2026}. Planet
  designations in the third column are followed by the original TOI
  candidate suffix in parentheses.}
\end{deluxetable*}

We identify 4 multi-planet systems containing a total of 8 validated
planets, of which 3 systems (6 planets) are entirely new.
Table~\ref{tab:multi} lists the system properties, period ratios, and
architectural classifications.

Three of the four retained VP pairs are compact (both orbital periods
below $10$~d), and two systems (TOI-5489 and TOI-786)
exhibit ``peas-in-a-pod'' architectures with sibling radii within
$\sim 20\%$, consistent with the patterns
identified in the Kepler multi-planet population
\citep{weiss2018, millholland2017}. Two systems are within $2\%$ of
low-order period commensurabilities: \toi{6729} lies $1.4\%$ from the
2:1 mean-motion resonance and \toi{786} lies $1.4\%$ from 3:1. These
near-resonant systems may merit transit-timing variation (TTV)
follow-up \citep{agol2005, holman2005}.

Multi-planet candidates validate at a higher rate than single-planet candidates:
individual candidates in multi-planet systems validate at $32.6\%$
compared to $16.4\%$ for single-planet candidates ($1.99\times$). These rates are
consistent with the expectation that multi-transiting systems are
overwhelmingly genuine \citep[][estimated a false-positive rate below
$1\%$ for Kepler multis]{lissauer2012}. However, multi-candidate
systems in our sample are also systematically brighter and have more
observed sectors, so the observed multiplicity boost should be viewed
as suggestive rather than as an isolated multiplicity prior.

We highlight the TOI-1806 system, where we validate two new planets
(TOI-1806~c and TOI-1806~d) alongside the previously confirmed \toi{1806.01},
expanding this system to three planets spanning periods from $0.50$~d
(TOI-1806~c, an ultra-short-period super-Earth) to $15.15$~d (the
confirmed outer planet, .01). The host star is a cool M dwarf
($T_\mathrm{eff} = 3272$~K); the full three-planet
architecture spans a period ratio of $30.2$ between the confirmed outer
planet and TOI-1806~c, while our two newly validated
planets alone (c and d, shown in Table~\ref{tab:multi}) span a
ratio of $16.33$.\footnote{The previously confirmed \toi{1806.01} is
not assigned a planet letter in the NASA Exoplanet Archive. We assign
the letters c (\toi{1806.03}, $P = 0.50$~d) and d (\toi{1806.02},
$P = 8.20$~d) to the planets validated here, in period order, leaving b
available for the earlier-discovered \toi{1806.01}.}

To the extent that the multiplicity boost reflects the intrinsic
low false-positive rate of multi-transit systems rather than
observational selection, it complements the imaging-based validation
pathway that dominates for isolated candidates.

\subsection{Notable Individual Targets}

The following highlights illustrate the range of planets in the
validated catalog. Rather than discuss every validated planet, we
briefly highlight targets that populate under-explored parameter space
and refer readers to Table~\ref{tab:validated} and the machine-readable
catalog for the complete sample.

\textbf{Ultra-short-period planets.} TOI-6662 b
($P = 0.294$~d, $R_p = 0.94~\rearth$) is the smallest and
shortest-period planet in our catalog: a sub-Earth on a $7.05$-hour
orbit around an M dwarf ($T_\mathrm{eff} = 3359$~K), with $\fpp$
effectively zero. It would not meet the VP threshold without the
contrast curve ($\fpp_\mathrm{noCC} = 0.026$).

\textbf{Imaging-dependent validation.} TOI-4324 b
($R_p = 1.23~\rearth$, $P = 6.25$~d) is a small planet around an M dwarf
whose validation depends on the contrast curve: without imaging it would
fall to possible-planet status ($\fpp_\mathrm{noCC} = 0.132$), while with
its contrast curve $\fpp < 10^{-8}$. The false-positive probability the
curve removes is almost entirely the bound-companion transiting-planet
scenario (STP); by excluding a companion bright enough to host the
transit, the imaging constraint drives it to negligible values. This
case illustrates that high-resolution imaging can be necessary for the
validation of small TESS planets in crowded fields within our
\triceratops-based analysis.

\textbf{Bright host targets.} Several VPs orbit stars bright enough for
precise radial-velocity follow-up. TOI-4307 b
($R_p = 1.39~\rearth$, $P = 32.7$~d), the validated candidate in that
system, orbits a bright late-F dwarf
($T_\mathrm{eff} = 6079$~K, $T_\mathrm{mag} = 6.64$) at
$\fpp < 10^{-6}$, making it a bright candidate for radial-velocity
follow-up. TOI-2540 b ($T_\mathrm{mag} = 8.27$, K-dwarf host,
$R_p = 2.29~\rearth$) is also among the brightest planets in the final
VP catalog and is a favorable target for RV mass constraints.
The machine-readable catalog provides the full set of magnitudes and
derived parameters; we discuss the JWST subset in Section~\ref{sec:jwst}
and Table~\ref{tab:jwst-targets}.

\section{Discussion}\label{sec:discussion}

\subsection{Implications for TESS Validation}

Our results demonstrate that, within the \triceratops\ framework,
contrast curves are a necessary component of the validation
infrastructure for the majority of small validated planets with
available contrast curves in our sample. Without contrast curve constraints, $72\%$ of our validated
planets with adopted contrast curves would fail the
$\fpp < 0.015$ threshold, reverting from validated planet to possible
planet status. The effect is strongly size-dependent: $100\%$ of
validated planets below $1.7~\rearth$ require the contrast curve to
achieve validation, compared with $33\%$ of sub-Saturns above
$4.0~\rearth$. This gradient reflects the underlying physics: smaller planets produce
shallower transits that are more easily reproduced by an eclipsing binary
or transiting planet on an unresolved companion, and the contrast curve
directly constrains the brightness of any such companion. As shown in
Section~\ref{sec:mechanism}, the probability the curve removes is carried
predominantly by unresolved bound companions rather than chance-aligned
background blends: the curve validates by ruling out a bound stellar
companion bright enough to host the transit signal, while the strongly
suppressed background-blend scenarios contribute little absolute
probability. Beyond being the predominant scenario impacted by the inclusion of
contrast curves, these bound companions are also known to have an impact on
planet occurrence rate, making their detection paramount for accurate planet
demographics studies \citep[e.g.,][]{moe2021binary, sullivan2026binary}.

Because the contrast curve validates through the bound-companion channel,
the properties of that channel set what high-resolution imaging should aim
for. These companions sit at modest brightness contrast, since a broad
mass-ratio distribution places a substantial fraction at comparable mass
\citep{duchene2013}, and their projected separations follow a broad
distribution peaking near tens of astronomical units \citep{raghavan2010,
duchene2013}, of which a typical contrast curve resolves only the outer
part. As a model estimate, integrating the \citet{raghavan2010} separation
distribution for a representative target (a $0.3''$ inner working angle at
$100$~pc), a factor-of-two improvement in angular resolution removes only
about a fifth of the unresolved companions and an order-of-magnitude
improvement about half. Because these companions lie predominantly at small
separations, the operative property for the bound channel is sensitivity to
faint companions close to the star, with Gaia astrometry independently
flagging a fraction of close unresolved companions \citep{belokurov2020};
which follow-up best achieves this for a given target depends on its
distance and host properties and is beyond the scope of this work.

The candidates imaging fails to validate include 43 that remain
bound-dominated even with a curve. For these the constraint does lower the
bound-companion probability, but the residual stays above the validation
threshold; moreover, the FPP is the sole barrier for only 14 of the 43, the
other 29 also failing the nearby-star (NFPP) or Gaia-binarity criteria,
which a contrast curve cannot address. Those 14 are the natural population
for which deeper follow-up on the bound channel might prove decisive.

Which contrast-curve property most efficiently suppresses the
bound-companion channel, whether inner working angle, close-separation
contrast, wavelength, or wide-separation depth, remains an open question
that our heterogeneous follow-up cannot isolate, since speckle,
adaptive-optics, and aperture-masking observations differ simultaneously in
wavelength, angular resolution, and contrast floor \citep{furlan2017,
kraus2016}. One published expectation is that redder bandpasses improve
sensitivity to cool, low-mass companions \citep{matson2018, furlan2017,
lester2021}; their relative importance must be calibrated empirically, for
instance with paired observations that vary one property at a time or
controlled re-runs in which the adopted curve is replaced while all other
inputs are held fixed.

This finding carries practical implications for the TESS validation
pipeline. The TESS Follow-up Observing Program (TFOP) Sub-Group 3 (SG3)
conducts high-resolution imaging for planet candidates
\citep{howell2021, ziegler2020, ziegler2021}, but observing
time is finite and must be prioritized. Among the targets validated in
this work, our results suggest the highest marginal return for
small planets and M-dwarf hosts: CC-dependence rises to
$100\%$ for super-Earth-sized and smaller planets, and M
dwarfs show the highest CC-dependence of any host spectral type
($83\%$; Section~\ref{sec:cc_impact}).
All else equal, sub-Saturn candidates may warrant lower imaging
priority when other follow-up resources are constrained, as their deeper
transits already provide substantial leverage against blended scenarios.

We note that our sample is subject to a selection effect: targets with
contrast curves available on ExoFOP \citep{christiansen2025} may represent systems that were
already considered promising by the community. The higher validation
rate among targets with contrast curves ($25.8\%$ vs.\ $6.7\%$) therefore
reflects both the constraining power of the imaging data and a
potential brightness or observability bias. The size dependence of the
CC-dependence rate, by contrast, reflects the physical mechanism of
Section~\ref{sec:cc_impact}: small planets produce shallow transits that
are more readily mimicked by blended eclipsing binaries, so contrast
curves are more often decisive at small radii.

\subsection{JWST Follow-up Targets}\label{sec:jwst}

We compute TSM following \citet{kempton2018}, using our MCMC-derived
planet radii, TIC~v8.2 $J$ magnitudes and stellar radii
\citep{stassun2019}, and planet
masses estimated from the \citet{chenkipping2017} mass--radius relation
adopted in that work. With this implementation, we identify 8
validated planets that exceed the broad
high-priority thresholds for transmission spectroscopy (TSM)
recommended in that work. Four of
the 8 are newly validated, and all 8 are contrast-curve-dependent:
they would not achieve validation without the imaging constraint. Using
the TSM size classes, 7 lie in the terrestrial bin
($R_p < 1.5~\rearth$) and 1 in the sub-Neptune bin
($1.5$--$2.75~\rearth$). We compute ESM only for the terrestrial
subset where the emission-spectroscopy threshold is
applicable ($R_p < 1.5~\rearth$).
Two terrestrial VPs exceed
$\mathrm{ESM} > 7.5$, one of which is newly validated in this catalog:
TOI-6662 b. Table~\ref{tab:jwst-targets}
summarizes the newly validated planets that meet these formal
follow-up criteria.

\begin{deluxetable*}{lllcccccc}
\tabletypesize{\scriptsize}
\tablecaption{Notable newly validated atmospheric follow-up candidates.
  The table lists newly validated planets that exceed either the broad
  \citet{kempton2018} TSM thresholds or the terrestrial ESM threshold
  ($R_p < 1.5~\rearth$ and $\mathrm{ESM} > 7.5$). All listed planets are
  contrast-curve-dependent under the no-contrast-curve validation test.
  \label{tab:jwst-targets}}
\tablehead{
  \colhead{Planet} &
  \colhead{TOI} &
  \colhead{Selection} &
  \colhead{$R_p$ ($\rearth$)} &
  \colhead{$P$ (d)} &
  \colhead{$T_\mathrm{eq}$ (K)} &
  \colhead{$M_p$ ($\mearth$)} &
  \colhead{TSM} &
  \colhead{ESM}
}
\startdata
TOI-6662 b & TOI-6662.01 & TSM+ESM & 0.94 & 0.29 & 1114 & 0.78 & 27.9 & 12.0 \\
TOI-2540 b & TOI-2540.01 & TSM & 2.29 & 12.72 & 477 & 5.87 & 104.5 & \nodata \\
TOI-4324 b & TOI-4324.01 & TSM & 1.23 & 6.25 & 439 & 2.02 & 10.7 & 1.5 \\
TOI-5961 b & TOI-5961.01 & TSM & 1.38 & 1.62 & 956 & 2.47 & 10.4 & 6.6 \\
\enddata
\tablecomments{TSM and ESM are computed following \citet{kempton2018}.
  Planet radii are from the robust time-domain MCMC fits, $J$ and $K$
  magnitudes and stellar radii are from TIC~v8.2 \citep{stassun2019},
  and masses are estimated from the \citet{chenkipping2017} mass--radius
  relation adopted by \citet{kempton2018}. Selection labels denote planets
  above the broad TSM threshold for their radius class (TSM), above the
  terrestrial ESM threshold (ESM), or both. ESM is reported only for
  terrestrial planets ($R_p < 1.5~\rearth$), where the \citet{kempton2018}
  threshold is applicable; \nodata\ indicates that the ESM threshold is not
  applied.}
\end{deluxetable*}

The top transmission spectroscopy target is TOI-2540 b, a
$2.29~\rearth$ sub-Neptune on a $12.72$-day orbit around a K dwarf
($T_\mathrm{mag} = 8.27$). Its TSM of about 105 places it above the
threshold for its size class ($\mathrm{TSM} > 90$
for $1.5$--$2.75~\rearth$). Without the contrast curve, the FPP for
this target rises from $0.009\%$ to $4.6\%$, illustrating a case where
a single high-resolution image makes the difference between a validated
JWST-accessible planet and a possible planet.

Among the newly validated terrestrial emission candidates, TOI-6662 b
has the highest ESM ($\mathrm{ESM} = 12.0$). It is an
ultra-short-period planet around a cool star and is
contrast-curve-dependent, so its inclusion among validated
emission-spectroscopy candidates also depends on the imaging constraint.

These results reinforce a broader point: within our \triceratops-based
analysis, the contrast curve infrastructure maintained by TFOP SG3
enables the validation of targets commonly prioritized for atmospheric
characterization
\citep[e.g.,][]{hord2024}.
These targets would benefit from radial velocity confirmation to refine
their mass estimates and atmospheric metric values.
\section{Conclusions}\label{sec:conclusions}

We applied \triceratops\ to 603 TESS Objects of Interest and
statistically validated 80 planets (64 new) spanning
$0.94$--$7.83~\rearth$ in radius and $0.29$--$47.0$~days in orbital
period, with host stars from mid-M to mid-F. The catalog includes
17 planets in the Neptunian desert ($P < 4$~d, $3$--$8~\rearth$) and
four multi-planet systems (three entirely new, 8 VPs total); no
giant planets ($R_p > 8~\rearth$) meet the FPP threshold.

The controlled removal experiment is the central result: among the 68
VPs with adopted contrast curves, $72\%$ would fail the
$\fpp < 0.015$ threshold if their imaging were withdrawn, rising to
$100\%$ for planets below $1.7~\rearth$. All 8 validated planets
exceeding the \citet{kempton2018} TSM thresholds for JWST transmission
spectroscopy are contrast-curve-dependent. The practical
implication is that, within our curated TESS sample and
\triceratops-based validation setup, the yield of small-planet
statistical validation is strongly limited by the availability of
high-resolution imaging.

The imaging-based mechanism alone, however, does not resolve all
competing false-positive channels: among non-validated candidates the
contrast curve sharply reduces, but does not eliminate, the
bound-companion transiting-planet (STP) scenario, as defined in the
\triceratops\ framework \citep{giacalone2021} (Section~\ref{sec:cc_impact}).
In the residual it persists at a level comparable to the
nearby-resolved-star (NFPP) channel that imaging also cannot address,
inside the small-separation regime where contrast curves provide limited
leverage.
Reducing this residual population, and turning validated planets
into confirmed planets with measured masses, will require
complementary radial-velocity and astrometric follow-up.

We provide a machine-readable table containing the full validated
planet catalog with per-target FPP, NFPP, and FPP computed without
contrast curve constraints ($\fpp_\mathrm{noCC}$). Reporting both FPP
values for every target with an adopted contrast curve
allows future studies to assess the contribution of imaging to any
individual validation and to re-evaluate classifications as improved
imaging or other follow-up becomes available.

\section*{Data Availability}

The tables in this article are provided in machine-readable form. These
comprise the 64 newly validated planets with their derived planetary
parameters, FPP, NFPP, and contrast-curve-removed FPP
($\fpp_\mathrm{noCC}$); the 16 previously published planets we recover,
with their FPP and $\fpp_\mathrm{noCC}$; the 141 excluded targets with
their exclusion reasons; and the contrast-curve provenance for each
validated planet with adopted high-resolution imaging. We additionally
provide the per-scenario false-positive probabilities, with and without
the contrast curve, for the 80 validated planets. The TESS light curves
are available from the Mikulski Archive for Space Telescopes (MAST), and
the high-resolution imaging products are available from ExoFOP
\citep{christiansen2025}.

\section*{Acknowledgments}

S.G. is supported by an NSF Astronomy and Astrophysics Postdoctoral Fellowship under award AST-2303922. Funding for K.B. was provided by the European Union (ERC AdG SUBSTELLAR, GA 101054354). A.C.K. gratefully acknowledges the financial support of UK Science and Technology Facilities. This work is partly supported by JSPS KAKENHI Grant Numbers JP24H00017, JP25K24620, JP26H01402, JP24K00689 and JP26K00755.

We gratefully acknowledge the staff at Lick Observatory for their work in
operating and maintaining the Shane telescope and the ShARCS instrument.

Based on observations obtained at the Hale Telescope, Palomar
Observatory, as part of a collaborative agreement between the Caltech
Optical Observatories and the Jet Propulsion Laboratory (operated by
Caltech for NASA).

Some of the data presented herein were obtained at Keck Observatory,
which is a private 501(c)3 non-profit organization operated as a
scientific partnership among the California Institute of Technology,
the University of California, and the National Aeronautics and Space
Administration. The Observatory was made possible by the generous
financial support of the W.~M. Keck Foundation. The authors wish to
recognize and acknowledge the very significant cultural role and
reverence that the summit of Maunakea has always had within the Native
Hawaiian community. We are most fortunate to have the opportunity to
conduct observations from this mountain.

Some of the observations in this paper made use of the High-Resolution
Imaging instruments 'Alopeke and Zorro. 'Alopeke and Zorro were funded by
the NASA Exoplanet Exploration Program and built at the NASA Ames Research
Center by Steve B. Howell, Nic Scott, Elliott P. Horch, and Emmett Quigley.
'Alopeke and Zorro were mounted on the Gemini North and South telescopes,
respectively, of the international Gemini Observatory, a program of NSF's
OIR Lab, which is managed by the Association of Universities for Research in
Astronomy (AURA) under a cooperative agreement with the National Science
Foundation on behalf of the Gemini partnership: the National Science
Foundation (United States), National Research Council (Canada), Agencia
Nacional de Investigaci\'{o}n y Desarrollo (Chile), Ministerio de Ciencia,
Tecnolog\'{i}a e Innovaci\'{o}n (Argentina), Minist\'{e}rio da Ci\^{e}ncia,
Tecnologia, Inova\c{c}\~{o}es e Comunica\c{c}\~{o}es (Brazil), and Korea
Astronomy and Space Science Institute (Republic of Korea).

This article is based on observations made with the MuSCAT2 instrument, developed by ABC, at Telescopio Carlos Sánchez operated on the island of Tenerife by the IAC in the Spanish Observatorio del Teide. This work is partly supported by JSPS KAKENHI Grant Numbers JP18H01265 and JP18H05439, and JST PRESTO Grant Number JPMJPR1775. We acknowledge financial support from the Agencia Estatal de Investigaci\'on of the Ministerio de Ciencia e Innovaci\'on MCIN/AEI/10.13039/501100011033 and the ERDF “A way of making Europe” through projects PID2021-125627OB-C32 and PID2024-158486OB-C32. This work is supported by the European Union (ERC AdvG SPEAR, GA 101200674). Views and opinions expressed are however those of the authors only and do not necessarily reflect those of the European Union or the European Research Council. Neither the European Union nor the granting authority can be held responsible for them.

In this study, the observational data obtained within the scope of project numbered 25ATUG100-3012 carried out using the TUG100 telescope at the TUG (T\"UBİTAK National Observatory, Antalya) site of the T\"urkiye National Observatories has been utilised, and we express our gratitude for the invaluable support provided by the T\"urkiye National Observatories, the observation team, and all its staff. We gratefully acknowledge the support by The Scientific and Technological Research Council of Turkey (T\"UB\.{I}TAK) with the project 126F041.

TRAPPIST, which collected data relevant to this analysis, is funded by the Belgian Fund for Scientific Research (Fond National de la Recherche Scientifique, FNRS) under the grant FRFC 2.5.594.09.F, with the participation of the Swiss National Science Foundation (SNF). MG and EJ are F.R.S.-FNRS Research Directors.

We acknowledge the support of the staff of the Xinglong 85cm and 80cm telescopes. This work was partially supported by National Astronomical Observatories, Chinese Academy of Sciences.

This research has made use of the Exoplanet Follow-up Observation Program
(ExoFOP; DOI: 10.26134/ExoFOP5) website, which is operated by the California
Institute of Technology, under contract with the National Aeronautics and
Space Administration under the Exoplanet Exploration Program.

This paper includes data collected with the TESS mission, obtained from the
MAST data archive at the Space Telescope Science Institute (STScI). Funding
for the TESS mission is provided by NASA's Science Mission Directorate.
STScI is operated by the Association of Universities for Research in
Astronomy, Inc., under NASA contract NAS5-26555. We acknowledge the use of
public TESS data from pipelines at the TESS Science Office and at the TESS
Science Processing Operations Center.

This work has made use of data from the European Space Agency (ESA) mission
Gaia (\url{https://www.cosmos.esa.int/gaia}), processed by the Gaia Data
Processing and Analysis Consortium (DPAC,
\url{https://www.cosmos.esa.int/web/gaia/dpac/consortium}). Funding for the
DPAC has been provided by national institutions, in particular the
institutions participating in the Gaia Multilateral Agreement.

The authors used Anthropic Claude Code with Claude Sonnet 4.6,
Claude Opus 4.6, Claude Opus 4.7, and Claude Opus 4.8, and OpenAI Codex with
GPT-5.4 and GPT-5.5, accessed between March and June 2026, to assist
with drafting and revising manuscript text, writing and reviewing analysis
code, organizing and checking computational workflows, generating and
checking figures and tables from the study data, and identifying candidate
references for author review. All references, calculations, scientific
interpretations, figures, tables, and manuscript text were reviewed and
verified by the authors, who take full responsibility for the accuracy,
integrity, and originality of the work.

\facilities{TESS, Gaia, Gemini:North, Gemini:South, Hale, Keck:II, Shane}
\software{%
  \triceratops\ \citep{giacalone2020},
  lightkurve \citep{lightkurve2018},
  SImMER \citep{savel2022},
  jaxoplanet \citep{jaxoplanet},
  NumPyro \citep{phan2019},
  JAX \citep{jax2018},
  juliet \citep{espinoza2019juliet},
  batman \citep{kreidberg2015},
  emcee \citep{foremanmackey2013},
  astropy \citep{astropy2022}%
}

\appendix

As supplementary material, we provide the individual \triceratops\ scenarios probabilities for each of the validated TOIs in Table~\ref{tab8}, which is available in full as a machine-readable table.

\begin{deluxetable}{cccc}\label{tab8}
\tablecaption{Individual scenario probabilities for validated TOIs}
\tablehead{
   \colhead{Row number} & \colhead{Units} & \colhead{Label} & \colhead{Description}
}
\startdata
1 & \nodata & TOI & TESS Object of Interest number \\
2 & \nodata & Planet\_designation & Assigned planet name \\
3 & \nodata & Has\_CC & Binary indicating whether the host star has an available contrast curve \\
4 & \nodata & DEB\_withCC & DEB scenario probability with contrast curve \\
5 & \nodata & DEBx2P\_withCC & DEBx2P scenario probability with contrast curve \\
6 & \nodata & BTP\_withCC & BTP scenario probability with contrast curve \\
7 & \nodata & BEB\_withCC & BEB scenario probability with contrast curve \\
8 & \nodata & BEBx2P\_withCC & BEBx2P scenario probability with contrast curve \\
9 & \nodata & DTP\_withCC & DTP scenario probability with contrast curve \\
10 & \nodata & EB\_withCC & EB scenario probability with contrast curve \\
11 & \nodata & EBx2P\_withCC & EBx2P scenario probability with contrast curve \\
12 & \nodata & NEB\_withCC & NEB scenario probability with contrast curve \\
13 & \nodata & NEBx2P\_withCC & NEBx2P scenario probability with contrast curve \\
14 & \nodata & NTP\_withCC & NTP scenario probability with contrast curve \\
15 & \nodata & PEB\_withCC & PEB scenario probability with contrast curve \\
16 & \nodata & PEBx2P\_withCC & PEBx2P scenario probability with contrast curve \\
17 & \nodata & PTP\_withCC & PTP scenario probability with contrast curve \\
18 & \nodata & SEB\_withCC & SEB scenario probability with contrast curve \\
19 & \nodata & SEBx2P\_withCC & SEBx2P scenario probability with contrast curve \\
20 & \nodata & STP\_withCC & STP scenario probability with contrast curve \\
21 & \nodata & TP\_withCC & TP scenario probability with contrast curve \\
22 & \nodata & DEB\_noCC & DEB scenario probability without contrast curve \\
23 & \nodata & DEBx2P\_noCC & DEBx2P scenario probability without contrast curve \\
24 & \nodata & BTP\_noCC & BTP scenario probability without contrast curve \\
25 & \nodata & BEB\_noCC & BEB scenario probability without contrast curve \\
26 & \nodata & BEBx2P\_noCC & BEBx2P scenario probability without contrast curve \\
27 & \nodata & DTP\_noCC & DTP scenario probability without contrast curve \\
28 & \nodata & EB\_noCC & EB scenario probability without contrast curve \\
29 & \nodata & EBx2P\_noCC & EBx2P scenario probability without contrast curve \\
30 & \nodata & NEB\_noCC & NEB scenario probability without contrast curve \\
31 & \nodata & NEBx2P\_noCC & NEBx2P scenario probability without contrast curve \\
32 & \nodata & NTP\_noCC & NTP scenario probability without contrast curve \\
33 & \nodata & PEB\_noCC & PEB scenario probability without contrast curve \\
34 & \nodata & PEBx2P\_noCC & PEBx2P scenario probability without contrast curve \\
35 & \nodata & PTP\_noCC & PTP scenario probability without contrast curve \\
36 & \nodata & SEB\_noCC & SEB scenario probability without contrast curve \\
37 & \nodata & SEBx2P\_noCC & SEBx2P scenario probability without contrast curve \\
38 & \nodata & STP\_noCC & STP scenario probability without contrast curve \\
39 & \nodata & TP\_noCC & TP scenario probability without contrast curve \\
\enddata
\tablecomments{Table 8 is published in its entirety in the electronic 
edition of the {\it Astronomical Journal}.  A portion is shown here 
for guidance regarding its form and content. Note that in the full table, blank entries indicate that the scenario was absent from the model for that target (e.g., stars with no resolved, nearby stars have no plausible NTP or NEB scenarios). We refer the reader to \citet{giacalone2021} for the definitions of the individual scenarios in the table.}
\end{deluxetable}

\bibliographystyle{aasjournalv7}
\bibliography{refs}

\end{document}